\documentclass[12pt]{article}
\usepackage{amsmath}
\usepackage{graphicx}
\begin{document}
\baselineskip 0.6cm

\def\simgt{\mathrel{\lower2.5pt\vbox{\lineskip=0pt\baselineskip=0pt
           \hbox{$>$}\hbox{$\sim$}}}}
\def\simlt{\mathrel{\lower2.5pt\vbox{\lineskip=0pt\baselineskip=0pt
           \hbox{$<$}\hbox{$\sim$}}}}

\begin{titlepage}

\begin{flushright}
UCB-PTH-09/31 \\
\end{flushright}

\vskip 1.7cm

\begin{center}

{\Large \bf 
A Finely-Predicted Higgs Boson Mass \\ \vspace{-0.1cm}
from \\ \vspace{0.1cm}
A Finely-Tuned Weak Scale
}

\vskip 0.8cm

{\large Lawrence J. Hall and Yasunori Nomura}

\vskip 0.4cm

{\it Berkeley Center for Theoretical Physics, Department of Physics, \\
     and Theoretical Physics Group, Lawrence Berkeley National Laboratory, \\
     University of California, Berkeley, CA 94720, USA} \\
{\it and} \\
{\it Institute for the Physics and Mathematics of the Universe, \\
     University of Tokyo, Kashiwa 277-8568, Japan} \\

\abstract{
If supersymmetry is broken directly to the Standard Model at energies 
not very far from the unified scale, the Higgs boson mass lies 
in the range $(128~\mbox{--}~141)~{\rm GeV}$.  The end points of 
this range are tightly determined.  Theories with the Higgs boson 
dominantly in a single supermultiplet predict a mass at the upper 
edge, $(141 \pm 2)~{\rm GeV}$, with the uncertainty dominated by 
the experimental errors on the top quark mass and the QCD coupling. 
This edge prediction is remarkably insensitive to the supersymmetry 
breaking scale and to supersymmetric threshold corrections so 
that, in a wide class of theories, the theoretical uncertainties 
are at the level of $\pm 0.4~{\rm GeV}$.  A reduction in the 
uncertainties from the top quark mass and QCD coupling to the 
level of $\pm 0.3~{\rm GeV}$ may be possible at future colliders, 
increasing the accuracy of the confrontation with theory from $1.4\%$ 
to $0.4\%$.  Verification of this prediction would provide strong 
evidence for supersymmetry, broken at a very high scale of $\approx 
10^{14 \pm 2}~{\rm GeV}$, and also for a Higgs boson that is elementary 
up to this high scale, implying fine-tuning of the Higgs mass parameter 
by $\approx 20~\mbox{--}~28$ orders of magnitude.  Currently, the only 
known explanation for such fine-tuning is the multiverse.
}

\end{center}
\end{titlepage}

\section{Overview}
\label{sec:overview}

The Standard Model (SM), taken to include neutrino masses, has reigned 
supreme for over three decades.  Despite strenuous efforts, at lepton 
and hadron colliders and from astrophysical observation, there is no hard 
evidence to contradict the Standard Model together with General Relativity 
(SM~$+$~GR) as the entire effective theory of nature up to extraordinarily 
high energies.

Over these decades, there have been many theoretical arguments for physics 
beyond the SM, with supersymmetry figuring very prominently and having 
two very different theoretical motivations:
\begin{itemize}
\item
String theory contains a quantum theory of gravity, and is the leading 
candidate theory for the unification of all the fundamental interactions. 
It requires supersymmetry in a spacetime with extra spatial dimensions, 
but leaves open the question of the size of supersymmetry breaking, 
which experiment allows to be anywhere in the range of the weak scale 
to the string scale.
\item
If supersymmetry breaking in the SM sector, $\tilde{m}$, is of order the 
weak scale, $v$, then the smallness of the weak scale relative to the 
Planck scale can be naturally understood.  In particular, a fine-tuning 
of the Higgs mass parameter to thirty orders of magnitude is avoided, 
and an elegant radiative mechanism for breaking of electroweak symmetry 
emerges.
\end{itemize}
Taken together, the theoretical motivation for supersymmetry is high, 
with the hope that superpartners are in reach of current hadron colliders.

Have experiments given any hint, positive or negative, on whether 
supersymmetric particles are at the weak scale?
\begin{itemize}
\item
Since the first experiments at LEP, it has become clear that the three 
SM gauge couplings unify more precisely if the theory is supersymmetric, 
with $\tilde{m}$ of order $v$.  Threshold corrections at the unified 
scale required for unification are fully an order of magnitude smaller 
with weak scale supersymmetry than without.  These corrections can arise 
from a mild non-degeneracy of one or two small multiplets at the unified 
scale with supersymmetry, but more multiplets or larger splittings are 
required without supersymmetry.
\item
The lightest weak scale superpartner can be stable, providing a Weakly 
Interacting Massive Particle (WIMP) candidate for Dark Matter (DM). 
It is intriguing that WIMPs, particles with order unity dimensionless 
couplings and order $v$ dimensionful couplings, lead to the observed 
abundance of DM, at least within a few orders of magnitude.
\item
A light Higgs boson, as expected in the simplest theories with weak 
scale supersymmetry, has not been found.  These theories now require 
a tuning of parameters, typically at the percent level, to reproduce 
the observed weak gauge boson masses.
\end{itemize}

\begin{figure}[t]
  \center{\includegraphics[scale=1.0]{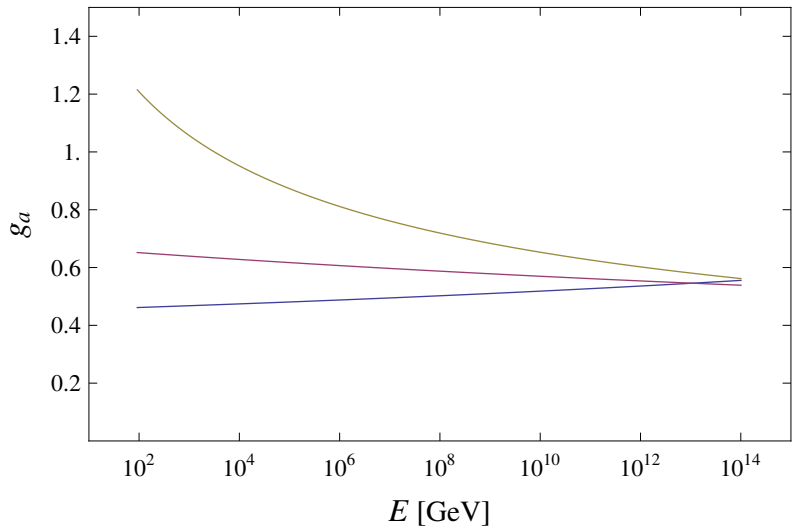}}
\caption{Evolution of the three gauge couplings, $g_a$ ($a=1,2,3$), 
 in the SM.  The $SU(5)$ normalization for the hypercharge gauge 
 coupling is taken.}
\label{fig:SM-gauge}
\end{figure}

In the first years after LEP, the first two items above provided 
a strong motivation for taking supersymmetry as the leading candidate 
for understanding the weak scale.  However, the absence of a light Higgs 
boson is certainly a problem for simple natural theories.  Furthermore, 
together with experimental bounds on superpartner masses, it pushes 
these theories into regions where the superpartner WIMP candidates 
are also unnatural.  This unease with weak scale supersymmetry is 
compounded by the lack of any signals of new flavor or $CP$ violation 
beyond the SM, such as $b \rightarrow s \gamma$, and by cosmological 
issues, such as the gravitino problem.  Over the years there were many 
opportunities for supersymmetry to become manifest, leaving us today 
with many reasons to question weak scale supersymmetry.  The single 
remaining success is gauge coupling unification, and while this is 
certainly significant, one wonders whether a decrease in the unified 
threshold corrections by an order of magnitude might be an unfortunate 
accident.  Even without supersymmetry, unification can occur, either 
by enhancing these threshold corrections or by certain matter surviving 
below the unified scale.  Indeed, the evolution of the gauge couplings 
in the SM shows evidence for unification~\cite{Georgi:1974yf}, as shown 
in Figure~\ref{fig:SM-gauge}, and precision unification requires only 
a small perturbation to this picture.

What, then, is the origin of the weak scale?  It has been suggested 
that the weak scale may result from anthropic, or environmental, 
selection~\cite{Agrawal:1997gf}.  In particular, if the Higgs mass 
parameter scans effectively in the multiverse, but not the Yukawa 
couplings, then the requirement of the stability of some complex nuclei 
requires that the weak scale be no more than a factor two larger than 
we measure~\cite{Agrawal:1997gf,Damour:2007uv}.  In this picture, 
most universes have weak interactions broken at a very high scale 
or by QCD dynamics, but they contain no complex nuclei and consequently 
no observers.  This view is often dismissed on the grounds that no 
evidence can be obtained for the multiverse, but this is incorrect. 
For example, consider Split Supersymmetry~\cite{ArkaniHamed:2004fb}: 
the weak scale is determined by environmental selection and supersymmetry 
is broken at some high scale $\tilde{m} \gg v$, but the fermionic 
superpartners are taken at the TeV scale to account for DM.  In this 
theory, collider measurements of the fermionic superpartner interactions 
could lead to a convincing determination of $\tilde{m}$ and demonstration 
that the Higgs field is elementary at the scale $\tilde{m}$.  This 
would imply a fine-tuning in the Higgs mass parameter of $1$ in 
$\tilde{m}^2/v^2$, which could be as large as $10^{20}$.  Fine-tuning 
that has no symmetry explanation is key evidence of the multiverse.

While environmental selection in the multiverse is speculative, 
we think it is the leading explanation for the order of magnitude 
of the cosmological dark energy~\cite{Weinberg:1987dv} providing 
the only understanding for $120$ orders of magnitude of fine-tuning. 
Indeed, we are greatly motivated by this result. Dark energy does 
not need any addition to the SM minimally coupled to gravity, and 
the prediction for the equation of state, $w = -1$, agrees well with 
current data, $w_{\rm obs} \simeq -1.0 \pm 0.1$~\cite{Komatsu:2008hk}. 
Of course, this requires a huge number of vacua, a landscape, that 
allows for sufficiently fine scanning of the cosmological constant, 
and it brings us back to the first theoretical motivation for 
supersymmetry, string theory, which is believed to have a landscape 
of sufficient size to allow the selection of both the cosmological 
constant and the weak scale~\cite{Bousso:2000xa}.

In this paper we assume that the weak scale is determined by environmental 
selection.  Where does that leave supersymmetry?  While the motivation 
from fine-tuning is gone, the motivation from string theory is 
strengthened, since the landscape has its origin in string theory. 
In seeking observational evidence for supersymmetry, the two key 
questions are then
\begin{itemize}
\item
What is $\tilde{m}$?
\item
Are there any non-SM particles near the weak scale?
\end{itemize}
We stress that, with the weak scale arising from environmental selection, 
we have lost the logical connection from naturalness between $\tilde{m}$ 
and $v$, and hence the expectation of superpartners at the weak scale.

The argument that some non-SM particles must survive to the weak 
scale, becoming WIMPs to account for DM, is not correct.  How is the 
strong $CP$ problem to be solved?  The small size of $CP$ violation in 
the strong interaction {\it must} be understood from conventional symmetry 
arguments; environmental selection cannot explain the smallness of the 
QCD angle, $\bar{\theta} \ll 1$, because there is no known catastrophic 
boundary involving $\bar{\theta}$.  Indeed, string theory is expected 
to contain a QCD axion, and therefore the Peccei-Quinn solution to the 
strong $CP$ problem~\cite{Peccei:1977hh}.  This leads to the expectation 
that axions~\cite{Weinberg:1977ma} are DM, with its density possibly 
determined by environmental requirements, removing any need for WIMP DM. 
Of course, there {\it could} be WIMP DM in addition to axion 
DM, but it is {\it not} necessary.

In this paper we therefore study the following simple framework: the 
supersymmetry breaking scale $\tilde{m}$ is very high, perhaps near 
the high energy cutoff of the field theory $M_*$, above which a string 
description becomes a necessity.  Below $\tilde{m}$, the effective 
theory is SM~$+$~GR.  Experimentally this sounds like a ``nightmare'' 
scenario, since the LHC may discover only the Higgs boson, with no 
hint of any physics beyond the SM.  This is, however, not true. 
We find that, although supersymmetry is broken at such high scales, 
a supersymmetric boundary condition on the Higgs quartic parameter 
is expected, leading to a narrow range for the Higgs boson mass of 
about $(128~\mbox{--}~141)~{\rm GeV}$.  Discovering a Higgs boson 
in this mass range would certainly be interesting, but it would be 
far more significant if the Higgs boson mass is close to the upper 
edge of this range.  This upper edge corresponds to the special 
situation that the Higgs boson resides dominantly in a single 
supermultiplet, and yields the prediction
\begin{equation}
  M_H = (141 \pm 2)~{\rm GeV}.
\label{eq:Higgs-pred-1}
\end{equation}
Remarkably, the largest contribution to the uncertainty results from 
the experimental errors on the top quark mass and the QCD coupling, 
which can be improved by future experiments to $\pm 0.3~{\rm GeV}$. 
The scenario can therefore be tested to high precision.

It is important that the prediction of Eq.~(\ref{eq:Higgs-pred-1}) 
does {\it not} depend sensitively on parameters that we cannot measure 
at low energies.  In a large class of theories, with $\tilde{m}$ 
ranging over a few orders of magnitude and with a variety of superpartner 
spectra, the theoretical uncertainties are extremely small, about 
$\pm 0.4~{\rm GeV}$ or less, reflecting both an infrared quasi-fixed 
point behavior of the Higgs quartic coupling and a reduced top Yukawa 
coupling at high energies.  Since the uncertainties arising from our 
lack of knowledge of the underlying high energy theory are so small, 
a measurement of this special value for the Higgs boson mass would 
provide strong evidence for the framework.

In fact, the prediction of Eq.~(\ref{eq:Higgs-pred-1}) survives even 
when the theory below $\tilde{m}$ is mildly extended beyond the SM. 
The conditions for such a precise prediction are that additional 
multiplets must make limited contributions to the beta functions 
of the SM gauge couplings, and that any new couplings to the Higgs 
boson must not be large.

A confirmation of the above Higgs mass prediction, together with the LHC 
finding no new physics beyond the SM, would provide significant evidence 
against our current paradigm and point to a very different picture of 
fundamental physics.  In fact, the observation of this single number 
would have many implications:
\begin{itemize}
\item[(i)]
Supersymmetry would be ``discovered,'' but with superpartners somewhere 
near $M_*$, rather than at the weak scale.  The discovery of supersymmetry 
would point to string theory, but the large breaking scale would radically 
change string compactification phenomenology.  All the ideas for new TeV 
physics---supersymmetry, technicolor, composite Higgs, and so on---would 
be replaced by the extension of the validity of the SM, perhaps augmented 
by a few small multiplets, up to very high energies.
\item[(ii)]
Axions provide the only compelling solution to the strong $CP$ problem, 
and hence axion DM would seem highly probable.  As the axion decay 
constant $f_A$ is expected to be very high, a pressing question becomes 
why the universe is not overclosed by axions.  This question has 
already been addressed: an environmental requirement on the density 
of DM may select the initial axion misalignment angle in our universe 
to be small~\cite{Linde:1987bx}.  WIMP DM, whether superpartners or 
not, would be unnecessary, although not excluded.
\item[(iii)]
The apparent success of supersymmetric gauge coupling unification would 
be seen to be an accident, that misled much of the field for two decades. 
The evolution of gauge couplings would still point to unification, as 
shown for the case of the SM in Figure~\ref{fig:SM-gauge}.  The SM alone 
requires larger unified threshold corrections, and leads to a lower, 
more uncertain, unification scale, $M_u \sim 10^{14 \pm 1}~{\rm GeV}$. 
Another possibility is that a few light multiplets additional to the 
SM lead to a precise unification, as in the case of a single vector-like 
lepton doublet near the weak scale.
\item[(iv)]
Most important, there would be a huge fine-tuning in the Higgs boson mass 
parameter of $20$ orders of magnitude or more.  The Higgs mass prediction 
would show that the Higgs boson is elementary up to very high energies, 
and there is no known symmetry mechanism that could tame the fine-tuning, 
given the high scale of supersymmetry breaking.  This would provide 
strong evidence that the electroweak symmetry breaking scale results 
from environmental selection.
\end{itemize}
To avoid these conclusions, one must either assume that the success of 
the Higgs mass prediction at the GeV level is an accident, or come up 
with an alternative understanding of the large amount of fine-tuning.

In the final section of the paper, we argue that certain other values 
of the Higgs boson mass could also demonstrate both an elementary 
Higgs boson to high scales and an absence of supersymmetry beneath 
the high scale, again providing evidence for environmental selection 
in the multiverse.

\section{A Supersymmetric Boundary Condition on {\boldmath $\lambda$}}
\label{sec:susy-bc}

If the SM becomes supersymmetric at scale $\tilde{m}$, then there is 
a boundary condition on the quartic Higgs coupling
\begin{equation}
  \lambda(\tilde{m}) 
  = \frac{g^2(\tilde{m}) + g'^2(\tilde{m})}{8} \cos^2\!2\beta,
\label{eq:lambda-bc_beta}
\end{equation}
where $g$ and $g'$ are the $SU(2)_L$ and $U(1)_Y$ gauge couplings, 
$g = g_2$ and $g' = \sqrt{3/5} g_1$.  The SM Higgs doublet is 
a combination of doublets of opposite hypercharge in the supersymmetric 
theory, described by a mixing angle $\beta$.  If $\tilde{m}$ is very 
large, does this boundary condition survive?  For example, suppose 
supersymmetry is broken by the highest component VEV, $F_X$, of a 
chiral superfield $X$, so that $\tilde{m} \sim F_X/M_*$.  In general, 
the K\"{a}hler potential includes the higher dimension operator 
$X^\dagger X (H^\dagger H)^2/M_*^4$ where $H$ is the Higgs superfield, 
so that the quartic coupling deviates from the supersymmetric 
boundary condition by an amount $\delta\lambda \sim F_X^2/M_*^4 
\sim \tilde{m}^2/M_*^2$.  With supersymmetry at the weak scale, 
$\tilde{m} \ll M_*$, so this correction is negligible; but for high 
scale supersymmetry breaking, does this correction destroy any Higgs 
mass prediction?

Many parameters, including $\tilde{m}$, are expected to vary in 
the multiverse.  High scale supersymmetry results if the landscape 
distribution for $\tilde{m}$ increases sufficiently rapidly at large 
$\tilde{m}$.  For a given value of $\tilde{m}$, we can determine 
whether a larger value is more probable by comparing whether the 
increase in probability from the $\tilde{m}$ distribution compensates 
for the more precise cancellation needed to keep $v$ below the 
environmental bound.  We expect that a larger $\tilde{m}$ is more 
probable if, at the value of $\tilde{m}$ under consideration, the 
$\tilde{m}$ distribution grows more rapidly than quadratically. 
As $\tilde{m}$ continues to grow, the distribution may become milder 
than quadratic, so that in typical universes observers find $\tilde{m} 
\ll M_*$.  However, in this case the form of the distribution introduces 
a new mass scale.  It seems more probable that the stronger peaking of 
the distribution persists all the way to near the cutoff $M_*$, so that 
typical observers find $\tilde{m}$ close to $M_*$.  This apparently 
destroys the boundary condition for $\lambda$ completely.  We argue 
below, however, that even in this case the supersymmetric boundary 
condition may well persist.

The new physics around the cutoff $M_*$ is likely to be accompanied 
by the compact spatial manifold that results from string theory. 
How large do we expect this new scale to be?  With $\tilde{m}$ near 
$M_*$, it is reasonable to assume that it is not far from the scale 
of SM gauge coupling unification, $M_u \approx 10^{14}~{\rm GeV}$. 
In this case the volume of the manifold is large, in units of the 
string scale, to account for the very large value of the Planck scale, 
$M_{\rm Pl} \approx 10^{18}~{\rm GeV}$.  There are two ways that such 
a setup may act to preserve the supersymmetric boundary condition. 
First, the strength of supersymmetry breaking may not really reach 
$M_*$.  For small supersymmetry breaking, an increase in $\tilde{m}$ 
is unlikely to affect the dynamics at $M_*$.  However, as $\tilde{m}$ 
approaches $M_*$, it may lead to a destabilization of the vacuum 
that yields the desired SM physics at low energy; $\tilde{m}$ may 
be prevented from reaching $M_*$ for an environmental reason.  The 
second possibility is that supersymmetry breaking is maximal but, 
because it is now occurring in a higher dimensional manifold, it 
is no longer true that it leads to sizable $\delta\lambda$.  Below 
we discuss ways in which the spatial properties of supersymmetry 
breaking can suppress $\delta\lambda$.

Supersymmetry breaking may either occur locally somewhere in the 
manifold, or it may be delocalized, as with Scherk-Schwarz or moduli 
breaking.  Local breaking of supersymmetry may typically occur far 
from the localization of the SM matter and Higgs sector.  In this case 
a non-local mediation mechanism is required and, given the large spatial 
separation, supersymmetry breaking in the SM Higgs sector is suppressed 
even if the local breaking of supersymmetry is maximal.  The non-local 
transmission may be by loops of quanta propagating in the bulk, which 
may include SM gauge fields.  The effects of tree-level transmissions 
are suppressed by the relevant volume factors; in particular, the 
gravity mediation contribution to $\delta\lambda$ is suppressed by 
$(M_*/M_{\rm Pl})^2$.  Once SM superpartners acquire mass, integrating 
them out gives loop threshold corrections to $\delta\lambda$.  These 
are computed in the next section and found to be small.

What if supersymmetry breaking is non-local?  In this case $\tilde{m}$ 
is determined by $\alpha/R$, where $\alpha$ ($\leq 1/2$) is an angle 
appearing in the compactification boundary conditions and $R$ is the 
size of the relevant extra dimension, which we take to be sufficiently 
larger than the cutoff scale for the classical spacetime picture to be 
valid.  Ignoring gravity, any tree-level corrections to $\delta\lambda$ 
are suppressed by powers of $\alpha/M_* R$.  There are loop threshold 
corrections to $\delta\lambda$ from integrating out superpartners and 
Kaluza-Klein (KK) excitations of SM particles.  The contributions 
from KK modes decouple if $\alpha$ is small and, as mentioned, the 
contributions from superpartners are small.  Even for $\alpha = 1/2$, 
the contribution from KK modes is loop suppressed.  The size of the 
gravity mediation contribution depends on the stabilization mechanism 
for the extra dimensions.  The correction to $\delta\lambda$, however, 
is suppressed by at least $(\alpha/M_* R)^2$ and typically much more.

Thus, even for maximal supersymmetry breaking, which likely leads to 
$\tilde{m}$ not far from $M_u$, the supersymmetric boundary condition 
for $\lambda$ may very well survive.  Indeed, the boundary condition 
is expected to be destroyed only in the very specific situation that 
supersymmetry breaking and the SM Higgs sector have coincident locations 
in the extra dimensions, and the supersymmetry breaking is maximal, 
with $F_X$ hard up against the cutoff.

\section{A Precise Prediction for the Higgs Boson Mass}
\label{sec:pred-higgs}

A prediction for the Higgs boson mass results from a supersymmetric 
boundary condition on the Higgs quartic coupling at $\tilde{m}$; 
however, the uncertainties might be very large.  Indeed, in the Minimal 
Supersymmetric Standard Model (MSSM) one-loop threshold corrections from 
top squark loops at $\tilde{m}$ lead to corrections to the Higgs boson 
mass as large as $\approx 40\%$.  For weak scale supersymmetry, collider 
measurements of superpartner properties could determine the threshold 
corrections, but this is clearly not possible for supersymmetry breaking 
at unified scales.  In this section we show that this naive expectation, 
of large uncertainties to the Higgs mass prediction from threshold 
corrections, is completely incorrect; rather, the largest uncertainties 
come from the experimental uncertainties on the top quark mass, $m_t$, 
and the QCD coupling, $\alpha_s$, which are already small and can be 
reduced by future precise measurements.

In section~\ref{subsec:Higgs-SM}, we compute the Higgs boson mass 
when the theory below $\tilde{m}$ is the SM, paying attention to 
possible threshold corrections from the scale $\tilde{m}$.  In 
section~\ref{subsec:Higgs-add}, we explore the sensitivity of the 
prediction to additional states with SM gauge interactions far 
below $\tilde{m}$, and in section~\ref{subsec:others} we discuss 
the relation to other work.

All figures and analytical results are obtained using two-loop 
renormalization group (RG) scaling of all couplings from $\tilde{m}$ 
to the weak scale, together with one-loop threshold corrections 
at the weak scale, including the one-loop effective potential 
for the Higgs field.  In addition, we include the two- and three-loop 
QCD threshold corrections in converting the top-quark pole mass 
to the $\overline{\rm MS}$ top Yukawa coupling, since they are 
anomalously large.  Experimental values of $m_t = 173.1 \pm 
1.3~{\rm GeV}$~\cite{Tevatron:2009ec} and $\alpha_s(M_Z) = 0.1176 
\pm 0.002$~\cite{Amsler:2008zzb} are used.

\subsection{SM below {\boldmath $\tilde{m}$}}
\label{subsec:Higgs-SM}

\begin{figure}[t]
  \center{\includegraphics[scale=1.0]{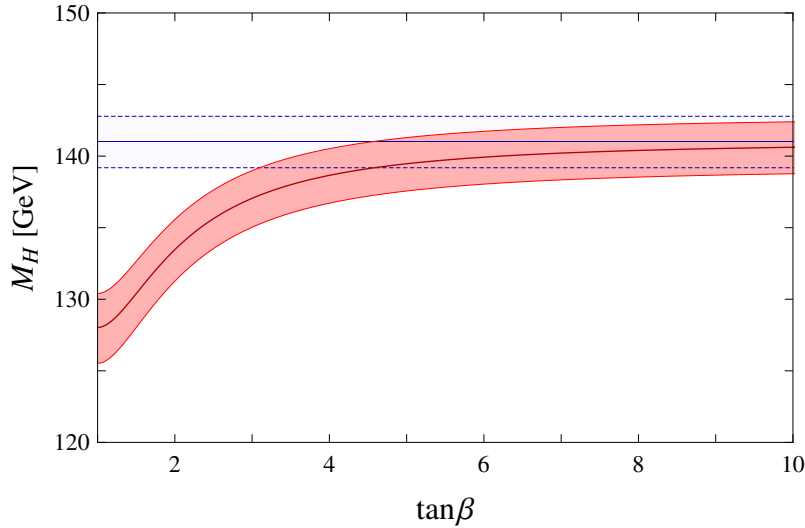}}
\caption{The Higgs mass prediction in the SM for theories where the 
 boundary condition for the quartic coupling at $\tilde{m}$ is given 
 by Eq.~(\ref{eq:lambda-bc_beta}), for fixed values of $\tilde{m} 
 = 10^{14}~{\rm GeV}$ and $\alpha_s(M_Z) = 0.1176$.  The solid red 
 curve gives the Higgs mass prediction for $m_t = 173.1~{\rm GeV}$, 
 while the shaded red band shows the uncertainty that arises from the 
 experimental uncertainty in the top quark mass of $\pm 1.3~{\rm GeV}$. 
 The horizontal blue lines show the corresponding asymptotes of the 
 prediction for large $\tan\beta$.  For $\tan\beta <1$, an identical 
 figure results provided the horizontal axis is labeled by $\cot\beta$.}
\label{fig:MH-tanb}
\end{figure}

In a general supersymmetric model, the SM Higgs doublet may be a 
combination of supersymmetric Higgs doublets having opposite hypercharge 
so that, before including threshold corrections, the boundary condition 
on the quartic coupling is given by Eq.~(\ref{eq:lambda-bc_beta}). 
The resulting prediction is actually a correlation between the Higgs 
boson mass and the parameter $\tan\beta$, as shown by the solid red 
curve in Figure~\ref{fig:MH-tanb}.
Remarkably, even as $\beta$ varies over all possible values, the Higgs 
mass lies in a narrow, high-scale supersymmetry, window of $\simeq 
(128~\mbox{--}~141)~{\rm GeV}$.  Furthermore, for large values of 
$\tan\beta$ the Higgs mass rapidly asymptotes to $\simeq 141~{\rm GeV}$, 
shown by the blue line, reaching $1~{\rm GeV}$ of this asymptote at 
$\tan\beta \simeq 6$.

As discussed in the next section, in many simple supersymmetric theories 
the parameter $\tan\beta$ is too large to be relevant or even does not 
exist, so that from now on we study the boundary condition
\begin{equation}
  \lambda(\tilde{m}) = \frac{g^2(\tilde{m}) + g'^2(\tilde{m})}{8} 
    \left\{ 1 + \delta(\tilde{m}) \right\},
\label{eq:lambda-bc}
\end{equation}
where $\delta$ includes all threshold corrections from the scale 
$\tilde{m}$, and is expected to be $\ll 1$ if $\tilde{m}$ is chosen 
close to the superparticle masses.  The effect of finite $\tan\beta$ 
can be included as a contribution to $\delta$
\begin{equation}
  \delta_{\beta} = - \frac{4}{\tan^2\!\beta} 
    + O\left( \frac{1}{\tan^4\!\beta} \right).
\label{eq:delta-beta}
\end{equation}

The Higgs mass prediction following from Eq.~(\ref{eq:lambda-bc}) takes 
the form $M_H = M_H(\tilde{m}, \delta(\tilde{m}))$, with both an explicit 
dependence on $\tilde{m}$ and an implicit one via $\delta$.  Since 
$\tilde{m}$ is an arbitrary matching scale, $M_H$ is independent of 
$\tilde{m}$: the explicit and implicit dependences cancel.  However, 
$M_H$ does depend on the spectrum of superpartners via the expression 
for $\delta$, with a typical sensitivity that can be estimated by 
studying the explicit dependence of $M_H$ on $\tilde{m}$, or equivalently 
on $\delta$.  As shown below, for a wide range of $\tilde{m}$ and $\delta$, 
these sensitivities of $M_H(\tilde{m}, \delta)$ are extremely mild.

\begin{figure}[t]
  \center{\includegraphics[scale=1.0]{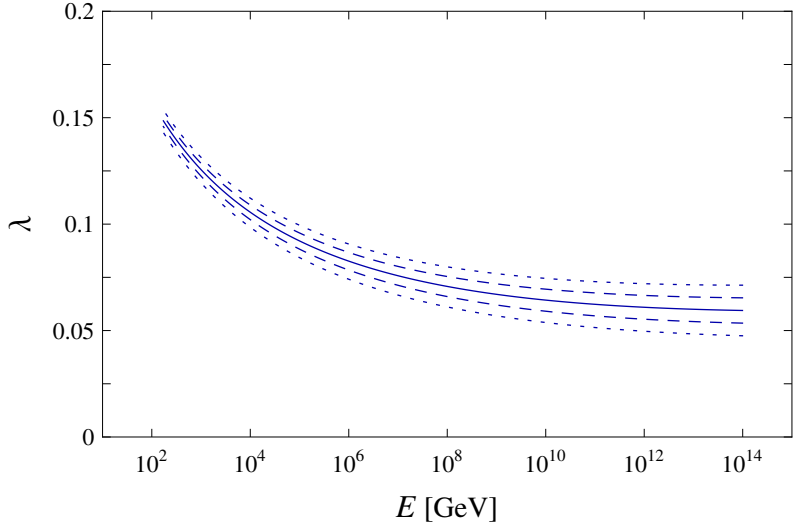}}
\caption{The evolution of the quartic coupling with energy $E$ in the SM 
 with the supersymmetric boundary condition of Eq.~(\ref{eq:lambda-bc}), 
 for fixed values of $\tilde{m} = 10^{14}~{\rm GeV}$, $m_t = 173.1~{\rm 
 GeV}$ and $\alpha_s(M_Z) = 0.1176$.  The solid curve is for $\delta 
 = 0$, while the long (short) dashed curves are for $\delta = \pm 0.1$ 
 ($\pm 0.2$).}
\label{fig:SM-lambda}
\end{figure}

In Figure~\ref{fig:SM-lambda}, we show the numerical solution for the 
running coupling $\lambda(E)$ as a function of energy $E$, for $\delta 
= 0$, $\pm 0.1$, and $\pm 0.2$ for $\tilde{m} = 10^{14}~{\rm GeV}$.
These curves show an important convergence property: the effects of the 
very large threshold corrections at $\tilde{m}$ are greatly reduced in 
the infrared.  The quartic coupling is being strongly attracted towards 
an infrared quasi-fixed point so that, at the weak scale, the fractional 
uncertainty in the coupling is reduced by about a factor of $6$.  This 
convergence of the infrared flow reduces the sensitivity of the Higgs 
boson mass to $\delta$
\begin{equation}
  \delta M_H = 0.10~{\rm GeV} \left( \frac{\delta}{0.01} \right),
\label{eq:delta-MH}
\end{equation}
where $\delta$ has been arbitrarily normalized to $0.01$.  Note that 
the attraction is not quite so strong as to erase the sensitivity of low 
energy measurements to the value of the supersymmetric boundary condition. 
This therefore still allows us to probe the existence of supersymmetry 
at high scales.

\begin{figure}[t]
  \center{\includegraphics[scale=1.0]{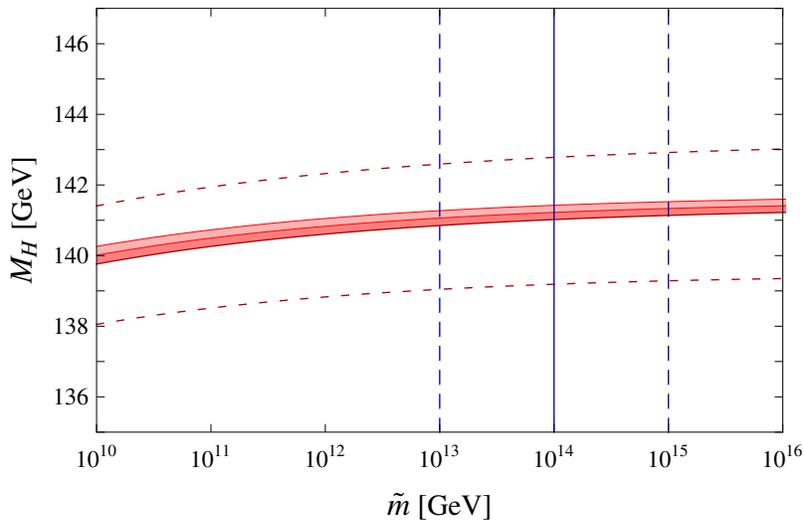}}
\caption{The explicit dependence of the Higgs mass prediction on 
 $\tilde{m}$ in the SM, with $\alpha_s(M_Z) = 0.1176$.  The narrow 
 red shaded region has $m_t = 173.1~{\rm GeV}$, with the three solid 
 curves corresponding to (from bottom) $\delta_s = 0$, $0.02$ and 
 $0.04$.  The upper (lower) dashed red curve shows the prediction 
 when the top quark mass is increased (decreased) by $1.3~{\rm GeV}$. 
 The vertical blue lines correspond to values of $\tilde{m}$ in the 
 region suggested by gauge coupling unification in the SM, $M_u 
 = 10^{14 \pm 1}~{\rm GeV}$.}
\label{fig:MH-mtilde}
\end{figure}

In Figures~\ref{fig:MH-tanb} and \ref{fig:SM-lambda} we have taken 
$\tilde{m} = 10^{14}~{\rm GeV}$ because, as we argued in the previous 
section, we expect supersymmetry breaking to be not far from the scale 
of unification, which from Figure~\ref{fig:SM-gauge} is seen to be of 
order $10^{14}~{\rm GeV}$.  However, Figure~\ref{fig:SM-gauge} also shows 
that $M_u$ has large uncertainties, and the superparticle masses may not 
be exactly at $M_u$.  An uncertainty in the Higgs boson mass induced by 
varying $\tilde{m}$ from $10^{14}~{\rm GeV}$, however, is extremely small
\begin{equation}
  \delta M_H = 0.14~{\rm GeV} 
    \left( \log_{10}\!\frac{\tilde{m}}{10^{14}~{\rm GeV}} \right),
\label{eq:delta-MH-2}
\end{equation}
as shown by the curves of Figure~\ref{fig:MH-mtilde} for a fixed value 
of $\delta$.
As $\tilde{m}$ increases above $10^{12}~{\rm GeV}$, it is apparent 
that the Higgs mass is remarkably insensitive to even large variations 
in $\tilde{m}$.  The Higgs mass changes by only $300~{\rm MeV}$ when 
$\tilde{m}$ is changed by two orders of magnitude.  The origin of this 
insensitivity can be seen from Figure~\ref{fig:SM-lambda}; the curves 
for $\lambda(E)$ have a very small gradient above $10^{10}~{\rm GeV}$ 
and, in addition, there is the convergence effect on scaling down to 
the weak scale.

We have seen that the predicted value of $M_H$ is rather insensitive 
to $\delta$ and $\tilde{m}$, but what definition of $\tilde{m}$ should 
we choose, and what is the value of $\delta$ with that $\tilde{m}$? 
A convenient choice for $\tilde{m}$ is such that the leading-log 
contributions to $\delta$ from the superpartners and the heavy Higgs 
doublet vanishes.  At the leading-log level, these threshold corrections 
are accounted for by choosing to match the full supersymmetric theory 
with the SM at an arbitrary scale $\tilde{m}$, and inserting a term 
in $\delta$ proportional to $\ln(m_{\tilde{\imath}}/\tilde{m})$ for 
each superpartner $\tilde{\imath}$ that is integrated out.  We can 
then make the choice of $\tilde{m} = \tilde{m}(m_{\tilde{\imath}})$ 
in such a way that the sum of these logarithmic terms vanishes.

In the appendix we compute the complete leading-log corrections to 
$\delta$ from all superpartners of SM particles and from the heavy 
Higgs doublet.  We find that these corrections vanish if we choose 
$\tilde{m}$ to be
\begin{equation}
  \tilde{m} \simeq \frac{m_\lambda^{1.6}}{m_{\tilde{t}}^{0.6}},
\label{eq:mtilde}
\end{equation}
where $m_\lambda$ and $m_{\tilde{t}}$ are the gaugino and top squark 
masses.  An important point is that, although $\tilde{m}$ defined in this 
way does not exactly coincide with any particular superparticle mass, 
it is in the vicinity of $m_\lambda$ and $m_{\tilde{t}}$, so we expect 
$\tilde{m}$ to be not far from $10^{14}~{\rm GeV}$.  Because the explicit 
dependence of $M_H$ on $\tilde{m}$ is very mild, this is enough to make 
a precise prediction for $M_H$.

This choice of $\tilde{m}$ completely eliminates the leading-log 
supersymmetric corrections.  The supersymmetric threshold correction, 
$\delta_s$, therefore contains only finite terms.  For example, the 
contribution from loops of top squarks at $\tilde{m}$ is
\begin{equation}
  \delta_s
  = \frac{3 y_t^4}{32\pi^2 \lambda} 
    \left( \frac{2 A_t^2}{m_{\tilde{t}}^2} 
    - \frac{A_t^4}{6 m_{\tilde{t}}^4} \right) 
  \simeq 0.007 \left( \frac{2 A_t^2}{m_{\tilde{t}}^2} 
    - \frac{A_t^4}{6 m_{\tilde{t}}^4} \right),
\label{eq:delta-s}
\end{equation}
where $A_t$ is the trilinear coupling of the top squarks to the Higgs 
boson.  The numerical size of this correction is much smaller than in 
the MSSM because, on scaling up to very large values of $\tilde{m}$, 
the top Yukawa coupling $y_t$ is reduced by about a factor two and 
the effect is proportional to the fourth power of $y_t$.  For $A_t 
= m_{\tilde{t}}$ ($3 m_{\tilde{t}}$), Eq.~(\ref{eq:delta-s}) gives 
$\delta_s \simeq 0.013$ ($0.031$), leading to an increase of $M_H$ 
of $0.1$~($0.3$)~${\rm GeV}$.  We expect that the size of the 
other finite supersymmetric threshold corrections, which we have 
not computed, does not exceed this order.  The effect of the 
supersymmetric correction is shown by the three solid red curves 
in Figure~\ref{fig:MH-mtilde} for $\delta_s =0, 0.02$ and $0.04$.

Other threshold corrections may be present, depending on the nature 
of the theory near $\tilde{m}$.  The Higgs mass prediction will be 
affected by any additional significant couplings of the Higgs boson 
at or below $\tilde{m}$.  Except for the top coupling, which we have 
already included, the Yukawa couplings to the quarks and charged leptons 
give negligible effects.  If neutrino masses are of Dirac type, then 
the neutrino Yukawa couplings are also very small and are irrelevant. 
However, for Majorana masses arising from the seesaw mechanism, there 
is the possibility of a correction to the Higgs mass if the right-handed 
neutrino mass, $M_R$, is less than $\tilde{m}$, in which case
\begin{equation}
  \delta_\nu = \frac{1}{8\pi^2} 
    \left( \frac{m_\nu^2 M_R^2}{\lambda v^4} - 2\frac{m_\nu M_R}{v^2} \right) 
    \ln\frac{\tilde{m}}{M_R} 
  \simeq 0.004\, \frac{M_R}{10^{14}~{\rm GeV}} 
    \left( 1.4\, \frac{M_R}{10^{14}~{\rm GeV}} -1 \right) 
    \ln\frac{\tilde{m}}{M_R},
\label{eq:delta-nu}
\end{equation}
where in the last expression we have taken $m_\nu = 0.05~{\rm eV}$, 
corresponding to the heaviest neutrino mass for the normal hierarchy 
spectrum.  The correction is small; $|\delta M_H| \simlt 0.1~{\rm GeV}$ 
for $M_R \approx 10^{14}~{\rm GeV}$ and completely negligible for 
$M_R \ll 10^{14}~{\rm GeV}$.  In the special case $\tilde{m} > M_R > 
10^{14}~{\rm GeV}$, the correction rapidly grows, giving $\delta M_H 
\approx 1~{\rm GeV}$ for $M_R = 5 \times 10^{14}~{\rm GeV}$, corresponding 
to a neutrino Yukawa coupling of $\approx 1$.  We stress that $\delta_\nu$ 
vanishes if right-handed neutrinos are above $\tilde{m}$.

Having discussed the threshold corrections at the scale $\tilde{m}$, we 
now turn to uncertainties that result from scaling between $\tilde{m}$ 
and $v$.  Indeed, at present the largest uncertainty in the Higgs mass 
prediction arises from the experimental uncertainties in $m_t$ and 
$\alpha_s$, which enter the RG equation for $\lambda$ at one and two 
loops, respectively.  The present $1.3~{\rm GeV}$ uncertainty in $m_t$ 
leads to a $1.8~{\rm GeV}$ uncertainty in the Higgs mass, as illustrated 
by the dashed curves of Figure~\ref{fig:MH-mtilde}.  A conservative estimate 
of the uncertainty in $\alpha_s$ is $\pm 0.002$~\cite{Amsler:2008zzb}, 
leading to $\delta M_H = \mp 1.0~{\rm GeV}$.  A recent analysis of all 
relevant data argues that the uncertainty in $\alpha_s$ is a factor 
three smaller~\cite{Bethke:2009jm}.

The final uncertainties arise from higher loop effects in RG scaling 
and in the top quark threshold correction.  First, the correction from 
three-loop QCD RG scaling decreases the Higgs mass by $0.2~{\rm GeV}$. 
We have not computed three-loop running from the top Yukawa coupling and 
$\lambda$, but do not expect these to be significantly larger than the 
three-loop QCD running.  Second, in going from the top quark pole mass 
to the $\overline{\rm MS}$ top Yukawa coupling, the QCD corrections 
reduce the Higgs mass by $11.9$, $2.7$ and $0.8~{\rm GeV}$ from one, 
two and three loops, respectively.  As the loop level is increased, the 
successive reductions of the corrections by $23\%$ and $30\%$ suggest 
that the four-loop effect will be of order $30\%$ of the three-loop 
correction, i.e. $0.24~{\rm GeV}$.  Hence, we arrive at a conservative 
estimate of the higher loop uncertainties in the Higgs mass prediction 
of $\pm 0.5~{\rm GeV}$.

Collecting these results leads to our final prediction for the Higgs 
boson mass in the SM
\begin{eqnarray}
  M_H &=& 141.0~{\rm GeV} 
    + 1.8~{\rm GeV} 
      \left( \frac{m_t - 173.1~{\rm GeV}}{1.3~{\rm GeV}} \right) 
    - 1.0~{\rm GeV} 
      \left( \frac{\alpha_s(M_Z)-0.1176}{0.002} \right)
\nonumber\\
  && {}
    + 0.14~{\rm GeV} 
      \left( \log_{10}\!\frac{\tilde{m}}{10^{14}~{\rm GeV}} \right) 
    + 0.10~{\rm GeV} 
      \left( \frac{\delta}{0.01} \right) 
    \pm 0.5~{\rm GeV},
\label{eq:Higgs-SUSY}
\end{eqnarray}
where $\delta = \delta_\beta + \delta_s + \delta_\nu + \cdots$. 
As explained above, $\delta_{\beta,\nu}$ may vanish, so that only 
$\delta_s$ is mandatory; thus we have chosen to scale $\delta$ by 
a numerical factor following from Eq.~(\ref{eq:delta-s}).  Our result 
shows that currently the largest uncertainties arise from the experimental 
error on $m_t$ and $\alpha_s$.  The uncertainties from high energy 
theories are very small, and only about $\pm 0.4~{\rm GeV}$ if we vary 
$\tilde{m}$ within two orders of magnitude from $10^{14}~{\rm GeV}$ 
and take $\delta \approx O(0.01~\mbox{--}~0.03)$.

How might this situation change in the future?  Studies at a future 
linear collider argue that the experimental uncertainties can be 
reduced to $\delta m_t \approx 100~{\rm MeV}$ (defined at short 
distances) and $\delta \alpha_s \approx 0.0012$~\cite{Abe:2001nnb}, 
which induce uncertainties in the Higgs mass prediction of 
$0.14~{\rm GeV}$ and $0.6~{\rm GeV}$, respectively.  The same 
study estimates the experimental uncertainty in the Higgs boson 
mass to be $\approx 100~{\rm MeV}$, so that the confrontation of 
the prediction with experiment is now limited by $0.6~{\rm GeV}$ 
from $\delta \alpha_s$.  With a Giga-Z sample, a linear collider 
may reach the much reduced uncertainty of $\delta \alpha_s \approx 
0.0005$~\cite{Bethke:2009jm}. Hence, in the future the prediction 
may take the form
\begin{eqnarray}
  M_H &=& (141.0 + \Delta)~{\rm GeV} 
    + 0.14~{\rm GeV} 
      \left( \frac{m_t - 173.1~{\rm GeV}}{0.1~{\rm GeV}} \right) 
    - 0.25~{\rm GeV} 
      \left( \frac{\alpha_s(M_Z)-0.1176}{0.0005} \right)
\nonumber\\
  && {}
    + 0.14~{\rm GeV} 
      \left( \log_{10}\!\frac{\tilde{m}}{10^{14}~{\rm GeV}} \right) 
    + 0.10~{\rm GeV} 
      \left( \frac{\delta}{0.01} \right),
\label{eq:Higgs-SUSY-future}
\end{eqnarray}
where experimental uncertainties are scaled by $1\sigma$ error bars. 
We have assumed sufficiently precise higher loop theoretical calculations, 
shifting the central value by $\Delta~{\rm GeV}$, with $|\Delta| 
\simlt 0.5$.

So far we have assumed that $\tilde{m}$ is sufficiently less than $M_u$ 
that the boundary condition does not receive tree-level modifications 
from the enlargement of the SM gauge group, or threshold corrections, 
$\delta_u$, from heavy states in the unified theory.  If the unified 
gauge group is $SU(5)$ there is no tree-level correction, but $\delta_u$ 
is model dependent.  Nevertheless, even when $\tilde{m}$ and $M_u$ 
are very close, it is reasonable for $\delta_u$ to be comparable to 
the threshold corrections required for gauge coupling unification, 
which are $6\%$ in $g^2$, leading to $\delta M_H \sim 0.6~{\rm GeV}$. 
If $\tilde{m} > M_u$ then the prediction will depend on the form of 
the RG equations in the non-supersymmetric unified theory between 
$\tilde{m}$ and $M_u$.  Although these are model dependent, it is 
worth stressing that the effect of any such corrections on the Higgs 
mass will be reduced due to the IR focusing effect of the quasi-fixed 
point in the SM RG equation for $\lambda$.

If the SM gauge group is enlarged at $\tilde{m}$ by $U(1)_\chi$ 
($\subset SO(10)/SU(5)$), there is a tree-level modification to 
the boundary condition
\begin{equation}
  \delta_\chi = \frac{4 g_\chi^2 q_\chi^2}{g^2 + g'^2} 
  \rightarrow \frac{1}{4},
\label{eq:delta-chi}
\end{equation}
where $g_\chi$ and $q_\chi$ are the $U(1)_\chi$ gauge coupling and 
charge of the Higgs field.  The last expression follows from taking 
$g_\chi$ equal to its unified value in $SO(10)$, giving $\delta M_H 
\simeq 2.4~{\rm GeV}$.  This correction becomes power suppressed as 
the $U(1)_\chi$ breaking scale is increased above $\tilde{m}$.

\subsection{Additional multiplets far below {\boldmath $\tilde{m}$}}
\label{subsec:Higgs-add}

The Higgs mass prediction of the previous section applied to the case 
that the effective theory below $\tilde{m}$ is the SM.  How does the 
prediction change as additions are made to the low energy theory? 
For example, if experiment is able to confirm this prediction to within 
$\pm 1~{\rm GeV}$, can we conclude that there are likely no other states 
at the weak scale beyond the SM?  We do not consider the possibility 
of adding light {\it scalars} below $\tilde{m}$; without an environmental 
selection, such scalars are extremely improbable in the landscape. 
Thus the scalar potential at the weak scale is that of the SM, with 
the physical Higgs boson mass depending on the single unknown parameter 
$\lambda(v)$.  How sensitive is this parameter to the addition of 
light fermions or gauge bosons?

The prediction does not survive if the SM gauge group is embedded in 
some larger group far below $\tilde{m}$.  For example, if the gauge 
group from $\tilde{m}$ to near the weak scale is $SU(4)_C \times SU(2)_L 
\times SU(2)_R$, then the central value of the prediction changes. 
On the other hand, an additional gauge sector has no effect on the 
prediction if none of the new fermions carry SM quantum numbers, 
and if the SM particles are neutral under the new gauge interaction. 
The prediction will change if the Higgs boson or top quark carries 
the new gauge interaction and the new gauge coupling is not small.

Without extending the SM gauge group, the addition of light fermions 
will significantly modify the Higgs boson mass prediction if
\begin{itemize}
\item
There are additional, large, renormalizable couplings involving either 
the Higgs boson or the top quark.
\item
The resulting additions to the beta function coefficients of the SM 
gauge interactions, $\varDelta b_a$, are significant.
\end{itemize}
While the former is model dependent, we can numerically study the latter 
in a rather model independent way.

\begin{figure}[t]
  \center{\includegraphics[scale=0.65]{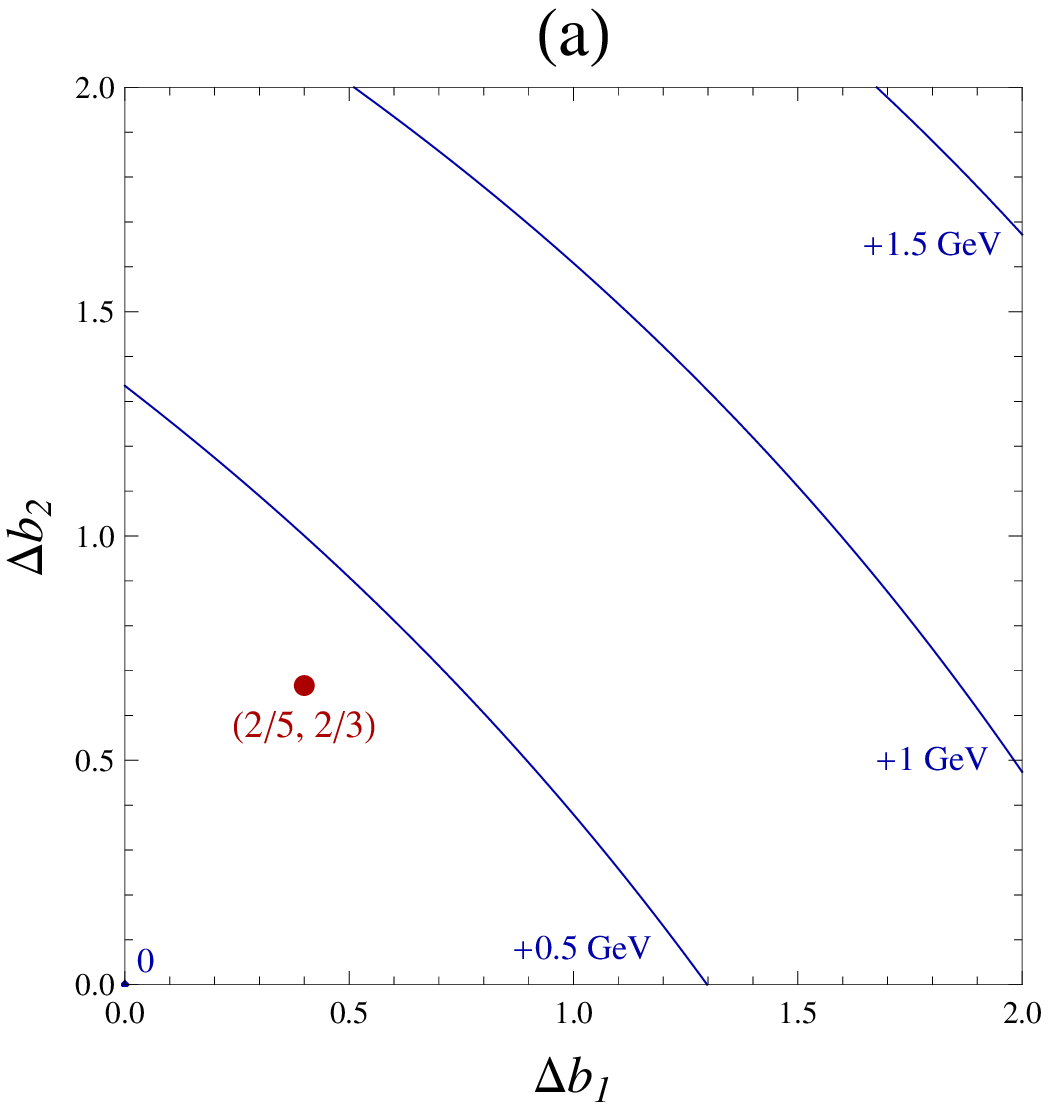}
  \hspace{1.5cm}
  \includegraphics[scale=0.65]{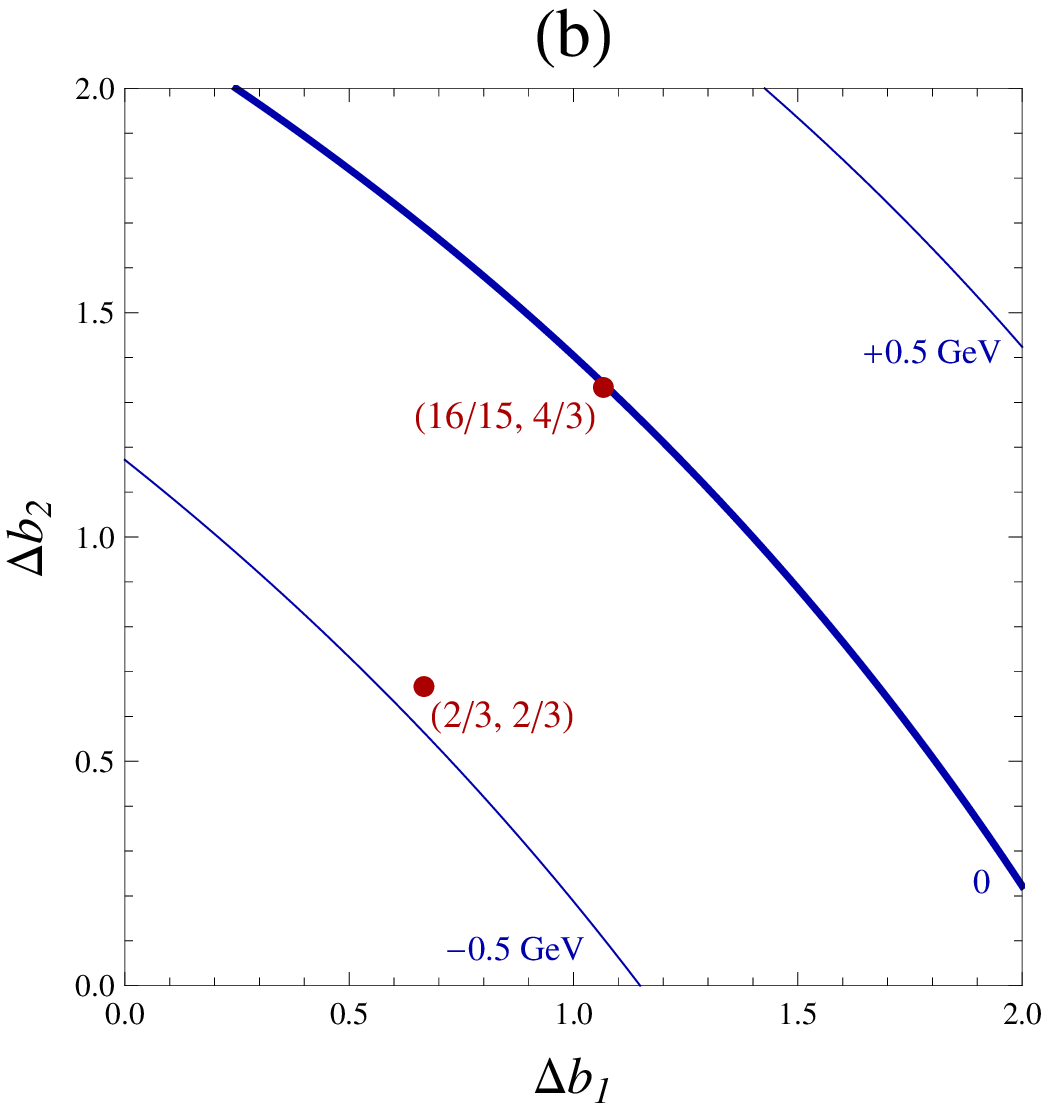}}
\caption{Contours of the shift in the Higgs mass prediction when 
 additional fermions of mass $1~{\rm TeV}$ are added to the SM. 
 These fermions contribute $\varDelta b_{1,2}$ to the $U(1)_Y$, $SU(2)_L$ 
 beta functions, but do not have significant Yukawa couplings to the 
 Higgs boson or top quark.  (a) None of the additional fermions are 
 colored.  The bold dot represents the addition of a single vector-like 
 lepton doublet.  (b) The only additional colored fermions are 
 a single vector-like triplet.  The bold dots represent the addition 
 of ${\bf 5} + {\bf \bar{5}}$ (lower) and ${\bf 5} + {\bf \bar{5}}$ 
 with a vector-like lepton doublet (upper).}
\label{fig:delta-b_a}
\end{figure}

If the additional fermions are all color singlets, contributions to 
$\varDelta b_{1,2}$ increase the Higgs mass, as shown by the contours 
of Figure~\ref{fig:delta-b_a}(a), where it is assumed that the mass 
of the additional fermions are $1~{\rm TeV}$.%
\footnote{If the masses are reduced to $100~\mbox{--}~200~{\rm GeV}$, 
 large additional corrections to the Higgs mass of order $1~{\rm GeV}$ 
 are induced, as the extra states give threshold corrections to the 
 values of the gauge couplings extracted from data.  These corrections 
 rapidly decouple as the mass of the extra states increases, and are 
 not included anywhere in this section.}
The addition of a single vector-like lepton doublet increases the Higgs 
mass by about $350~{\rm MeV}$, and is marked with a dot.  Note that 
$\varDelta b_2$ is quantized in units of $2/3$.  In order for the Higgs 
boson mass to stay within $1~{\rm GeV}$ of our prediction, only four 
additions with non-trivial $SU(2)_L$ are possible: one, two, or three 
vector-like doublets or one weak triplet.  The case of one vector-like 
lepton doublet, shown by the dot in Figure~\ref{fig:delta-b_a}(a), is 
particularly important, since it leads to gauge coupling unification 
that is as precise as for weak scale supersymmetry.

The most general theory with a single vector-like lepton doublet 
$(L, L^c)$, with no singlets, is described by the Lagrangian
\begin{equation}
  {\cal L} = {\cal L}_{\rm SM} + m L L^c + y L e h^\dagger.
\label{eq:L1vecdoub}
\end{equation}
The new Yukawa coupling ensures that the heavy lepton is unstable, 
which is crucial since otherwise the theory is excluded by limits on 
the direct detection of DM.  The charged and neutral heavy leptons, 
$L_E$ and $L_N$, will be pair produced at colliders, and each decays 
to an electroweak boson and a lepton $L_E \rightarrow (h,Z)(e,\mu,\tau),\, 
W \nu$ and $L_N \rightarrow W(e,\mu,\tau),\, (h,Z)\nu$.

An alternative possibility is that the vector-like lepton mixes with 
a neutral Majorana fermion so that, if the additional fermions are 
odd under some parity, the lightest neutral mass eigenstate is stable 
and, since it is Majorana, evades the DM direct detection limits. 
Indeed, these states result if the Higgsinos of the MSSM together 
with the bino or some other singlet fermion have masses far below 
$\tilde{m}$~\cite{ArkaniHamed:2005yv}.  In this case, new Yukawa 
interactions coupling the Higgs boson to the additional fermions may 
be present.  In this theory, a Higgs mass prediction follows from a 
supersymmetric boundary condition on the quartic coupling~\cite{EGHKN}, 
and depends on the size of the additional Yukawa interactions.

Adding colored fermions at the weak scale rapidly alters the Higgs 
mass prediction.  For example, a single vector-like color triplet 
without electroweak quantum numbers reduces the Higgs mass prediction 
by about $1~{\rm GeV}$.  In Figure~\ref{fig:delta-b_a}(b) we show 
contours of the change in the Higgs mass prediction for the case of 
a single vector-like color triplet when there are also contributions 
to $\varDelta b_{1,2}$, coming from the colored triplet itself or 
from additional electroweak states.  Two simple theories are shown 
by dots; one has states corresponding to $SU(5)$ multiplets ${\bf 5} 
+ {\bf \bar{5}}$, and the other has a further vector-like lepton doublet. 
This latter case has high precision gauge coupling unification and 
a Higgs mass prediction very close to the SM.  If accessible, the 
colored triplet, $D$, would be pair produced at the Tevatron or the 
LHC, with each decaying as $D \rightarrow (h,Z)(d,s,b),\, W(u,c,t)$ 
via the Yukawa interaction $qDh^\dagger$.  If $L$ mix with a singlet, 
the lightest state can be stable and contribute to DM.  However, the 
colored state $D$ must still decay via $qDh^\dagger$, since if this 
interaction is absent $D$ can decay only via dimension six operators 
and is cosmologically stable.

Figure~\ref{fig:delta-b_a} shows that only a very few weak-scale 
multiplets with small SM charges can be added to the theory if the 
Higgs mass prediction is to survive at the $\pm 1~{\rm GeV}$ level. 
Another possibility is to add multiplets at some scale $m$ intermediate 
between $v$ and $\tilde{m}$.  In the case that these states are 
non-colored, since the electroweak gauge couplings evolve slowly, 
Figure~\ref{fig:delta-b_a}(a) is still approximately correct providing 
the axis labeling is changed from $\varDelta b_a$ to $\varDelta b_a 
(\ln(\tilde{m}/m) / \ln(\tilde{m}/v))$.  Twice as many multiplets 
can be placed at $\sqrt{\tilde{m}v}$ compared to $v$.  Adding colored 
states at $v$ had a large effect on the Higgs mass because, although 
the effect is two loop via the effect on the top Yukawa, the QCD 
coupling is large at the weak scale.  By contrast, on adding states 
at intermediate scales, such as $\sqrt{\tilde{m}v}$, the change in the 
Higgs mass is dominated by $\varDelta b_{1,2}$ which gives an effect 
at one loop, rather than the two-loop effect from $\varDelta b_3$.

To conclude, experimental confirmation of the Higgs mass prediction 
of Eq.~(\ref{eq:Higgs-SUSY}), to an accuracy of $1~{\rm GeV}$, removes 
almost all alternatives to the SM at the TeV scale.  The addition of 
a vector-like lepton doublet remains as an interesting possibility.

\subsection{Relation to other work}
\label{subsec:others}

The theories illustrated by Figure~\ref{fig:delta-b_a} give a mild 
perturbation of less than $\pm 1~{\rm GeV}$ about the SM Higgs mass 
prediction.  The case of Split Supersymmetry~\cite{ArkaniHamed:2004fb} 
cannot be considered as a mild perturbation.  Indeed Split Supersymmetry 
is taken to include a very wide ranges of $\tilde{m}$ and $\tan\beta$, 
so that the Higgs mass can range from the present experimental limit 
of $114~{\rm GeV}$ up to about $155~{\rm GeV}$~\cite{Binger:2004nn}. 
Taking $\tilde{m}$ very high does not yield a central value 
close to the SM prediction: the light gluino contribution to 
$\varDelta b_3$ alone would decrease the Higgs mass prediction 
by about $3~\mbox{--}~5~{\rm GeV}$, but much more important are 
the new Yukawa couplings involving the Higgs boson, which increase 
the Higgs mass by $13$ to $19~{\rm GeV}$ depending on $\tilde{m}$ 
and $\tan\beta$.  For Split Supersymmetry with large $\tan\beta$ 
and $\tilde{m} \sim M_u \sim 10^{16}~{\rm GeV}$, a precise prediction 
for the Higgs mass emerges
\begin{equation}
  M_{H_{\rm split}} \simeq 154~{\rm GeV}.
\label{eq:MH-split}
\end{equation}
The precision of this special value of the Higgs mass within 
Split Supersymmetry can be defended at a level similar to that of 
Eq.~(\ref{eq:Higgs-SUSY}) for the SM.  Indeed, the threshold corrections 
involving electroweak gauginos are now at the weak scale, and could 
potentially be determined by measuring the electroweak gaugino 
masses and couplings.

Motivated by Split Supersymmetry, several groups have investigated 
supersymmetry breaking at a high scale, including models with 
supersymmetry breaking at a Peccei-Quinn breaking scale of 
$10^{11}~{\rm GeV}$~\cite{Barger:2004sf} and models with gauge 
coupling unification at $10^{16\mbox{--}17}~{\rm GeV}$ via non-$SU(5)$ 
hypercharge normalization~\cite{Barger:2005gn}.  In these models, 
a supersymmetric boundary condition on the quartic coupling yields 
a Higgs mass prediction and, for large values of $\tan\beta$ and 
taking account different values of the top quark mass, these 
predictions are not far from our central value of $141~{\rm GeV}$. 
This is a reflection of the remarkable insensitivity of the Higgs 
mass to variations in the unified scale and threshold corrections, 
as given in Eqs.~(\ref{eq:delta-MH}) and (\ref{eq:delta-MH-2}). 
Indeed, it will be difficult to use the Higgs mass prediction 
to distinguish between these schemes---for example, changing the 
unification scale from $10^{14}~{\rm GeV}$ to $10^{16}~{\rm GeV}$ 
changes the Higgs mass by less than $0.3~{\rm GeV}$.  Furthermore, 
the supersymmetric boundary condition on the Higgs quartic coupling 
does not depend on the Kac-Moody level relevant for gauge coupling 
unification.  On the other hand, the Higgs mass decreases significantly 
at low values of $\tan\beta$, as shown in Figure~\ref{fig:MH-tanb}, 
so that there is sensitivity to models that predict particular 
low values of $\tan\beta$~\cite{Gogoladze:2006ps}.

\section{Theories with High Scale Supersymmetry Breaking}
\label{sec:extra-dim}

We have explored the consequences of taking the SM as the correct 
effective theory up to some very high scale of supersymmetry 
breaking $\tilde{m} \sim M_u$, where the unification scale $M_u \sim 
10^{14 \pm 1}~{\rm GeV}$, as illustrated in Figure~\ref{fig:SM-gauge}. 
What is the new physics that emerges at this scale?  Since supersymmetry 
and the multiverse are both motivated by string theory, it is plausible 
that the higher dimensions of space are being encountered.  This offers 
the elegant possibility that breaking of both unified gauge symmetry 
and supersymmetry are associated with these extra dimensions; in 
particular, the unified gauge symmetry may be broken intrinsically 
by the compactification.  While a solution to the doublet-triplet 
splitting problem is no longer needed, such a framework has many 
appealing phenomenological features:
\begin{itemize}
\item
Proton stability is naturally accounted for, without the need for imposing 
any additional symmetries.  Since supersymmetry is broken at the high 
scale, there is no need to impose $R$ parity to avoid proton decay at 
dimension 4.  Indeed, proton stability is automatic at both dimension 
4 and 5.  With four-dimensional (4D) unification at $10^{14}~{\rm GeV}$, 
proton decay from gauge-mediated dimension 6 operators are 
disastrous, but this is easily avoided in higher dimensional 
theories~\cite{Hall:2001pg}.
\item
In simple theories the boundary conditions in extra dimensions, which 
involve very few parameters, can break both unified gauge symmetry and 
supersymmetry.  This gives simple KK towers of superpartners and unified 
states, allowing the calculation of threshold corrections to both gauge 
coupling unification and the Higgs quartic coupling.
\item
The requirement of two independent Higgs fields is removed: although 
different states of the supersymmetric theory couple to up and down 
quark sectors, these states may be part of the same supermultiplet 
in higher dimensions~\cite{Barbieri:2000vh}.
\item
The Higgs boson can be a slepton, allowing a unification of the matter 
and Higgs sectors of the SM.  This is not possible with low energy 
supersymmetry because of the masses and interactions that accompany 
the associated $R$ parity violation, but these constraints decouple 
as the scale of supersymmetry is raised.
\end{itemize}

In section~\ref{subsec:PQ} we show that, in theories where the boundary 
condition takes the form of Eq.~(\ref{eq:lambda-bc_beta}), an approximate 
symmetry, whether originating in four or more dimensions, leads 
to a sufficiently large $\tan\beta$ that the precise Higgs mass 
prediction of Fig.~\ref{fig:MH-mtilde} applies, with a very small 
correction from $\delta_\beta$ of Eq.~(\ref{eq:delta-beta}).  In 
section~\ref{subsec:single-H} we present a new, distinct class of 
theories which is particularly interesting in the context of high 
scale supersymmetry breaking.  In these theories, $\tan\beta$ does not 
exist and the boundary condition is given by Eq.~(\ref{eq:lambda-bc}). 
Although the Higgs boson mass in these theories can receive somewhat larger 
uncertainties than the ones discussed in section~\ref{subsec:Higgs-SM}, 
they are still at the level of a GeV.

\subsection{An approximate Peccei-Quinn symmetry}
\label{subsec:PQ}

In the case that the supersymmetric theory at $\tilde{m}$ is 4D, or 
that the two Higgs doublets of the supersymmetric theory, $h_{u,d}$, 
arise from different supermultiplets of a higher dimensional theory, 
the SM Higgs doublet is a linear combination of $h_{u,d}$
\begin{equation}
  h = h_u \sin\beta + h_d^\dagger \cos\beta.
\label{eq:beta-def}
\end{equation}
The boundary condition on the SM Higgs quartic coupling is then given 
by Eq.~(\ref{eq:lambda-bc_beta}) and depends on the mixing angle 
$\beta$.  However, for $\tan\beta \simgt 10$ the Higgs boson mass 
becomes very insensitive to $\beta$, varying by less than $0.4~{\rm GeV}$. 
A mechanism for large $\tan\beta$ can therefore lead to a very tight 
prediction for the Higgs boson mass.

If the theory possesses an approximate Peccei-Quinn symmetry, then the 
Higgsino mass parameter is suppressed, $\mu \sim \epsilon \tilde{m}$, and 
the mass matrix for the Higgs doublets $h_{u,d}$ takes the generic form
\begin{equation}
  \left( \begin{array}{cc}
    h_u^\dagger & h_d 
  \end{array} \right) 
  \left( \begin{array}{cc}
    \tilde{m}_2^2 & \epsilon \tilde{m}_3^2 \\
    \epsilon \tilde{m}_3^2 & \tilde{m}_1^2 
  \end{array} \right) 
  \left( \begin{array}{c}
    h_u \\ h_d^\dagger 
  \end{array} \right),
\label{eq:H-matrix}
\end{equation}
where $\epsilon$ is the small symmetry breaking parameter.  The parameters 
$\tilde{m}_{1,2,3}^2$ are typically of order $\tilde{m}^2$ and scan 
independently in the multiverse.  Given that environmental selection 
requires one eigenvalue of the above matrix to be of order $v^2$, what 
is the most probable value of $\tan\beta$ we observe?  In particular, 
is it more probable to have the determinant nearly vanish by having 
$\tilde{m}_{1,2}^2$ both suppressed by $\epsilon$, giving $\tan\beta 
\approx 1$, or by having one of them suppressed by $\epsilon^2$, so 
that $\tan\beta \approx 1/\epsilon$?  (We ignore the possibility of 
$\tan\beta \approx \epsilon$ since this is experimentally disfavored.) 
It turns out that the case of $\tan\beta \approx 1$ is less probable by 
a factor of $\epsilon$, since it implies that the heavier mass-squared 
eigenvalue is of order $\epsilon \tilde{m}^2$, requiring extra fine-tuning 
beyond that necessary to obtain the weak scale.  Hence, the approximate 
symmetry leads to the expectation
\begin{equation}
  \tan\beta \approx \frac{1}{\epsilon}.
\label{eq:large_tan-b}
\end{equation}

How small might $\epsilon$ be?  With dimensionless couplings of order 
unity, the bottom to top quark mass ratio takes the form
\begin{equation}
  \frac{m_b}{m_t} \approx \epsilon + c \frac{\tilde{m}}{M_*},
\label{eq:b-over-t}
\end{equation}
where the first term arises from the $b$ quark Yukawa coupling while 
the second term represents a possible contribution from higher dimension 
operators $[c(QD + LE)H_u^\dagger X^\dagger/M_*^2]_{\theta^4}$, where 
$c \ll 1$ or $\tilde{m} \ll M_*$ to preserve the boundary condition 
on $\lambda$, as discussed in section~\ref{sec:susy-bc}.  Thus the 
approximate Peccei-Quinn symmetry leads to an understanding of the 
small $m_b/m_t$ ratio for any
\begin{equation}
  \epsilon \simlt \frac{m_b}{m_t}.
\label{eq:epsilon}
\end{equation}
Conservatively, taking the upper limit to be $0.1$ leads to a contribution 
from $\delta_\beta$ to the Higgs boson mass of only $-0.4~{\rm GeV}$ and, 
for most values of $\epsilon$ that lead to an understanding for $m_b/m_t$, 
the contribution from $\delta_\beta$ is negligible.  Indeed, it is 
interesting to note that $\epsilon$ may be extremely small so that, 
for all practical purposes, $h = h_u$ and the $b$ quark mass originates 
entirely from the higher dimension operator.  In this case the 
Higgsino becomes light, and may be the vector-like lepton doublet 
of Eq.~(\ref{eq:L1vecdoub}).

The Peccei-Quinn symmetry described here may be responsible for the 
solution to the strong $CP$ problem, in which case we expect $\epsilon 
\sim f_a/\tilde{m}$, where $f_a$ is the axion decay constant, the 
scale at which the Peccei-Quinn symmetry is spontaneously broken. 
For example, this could result from a 4D superpotential interaction 
of the type $[S H_u H_d]_{\theta^2}$, with order unity coupling and 
the scalar component of $S$ acquiring a VEV of size $f_a$.  This would 
lead to $\mu \sim f_a$ as well as the suppressed Peccei-Quinn breaking 
mass in Eq.~(\ref{eq:H-matrix}).  With $f_a \sim 10^{12}~{\rm GeV}$ 
and $\tilde{m} \sim 10^{14}~{\rm GeV}$, one expects $\tan\beta \sim 
1/\epsilon \sim 10^2$, so that the correction to the Higgs mass 
prediction from $\delta_\beta$ is negligible.

In theories with extra spatial dimensions, the Higgs fields $h_{u,d}$ 
have profiles in the bulk, and the small parameter $\epsilon$ may 
result from a small overlap of the wavefunctions for $h_u$ and $h_d$. 
In this case, there is no need to impose an approximate symmetry 
on the higher dimensional theory; rather, it emerges in the 4D theory 
as a result of locality in the higher dimensions.  This origin for the 
small off-diagonal term in Eq.~(\ref{eq:H-matrix}) is somewhat general; 
no matter how many extra dimensions, a small $\epsilon$ results 
providing $h_u$ and $h_d$ profiles are peaked in differing locations. 
Strong peaking of the wavefunctions might arise, for example, from 
higher dimensional mass terms or from localizations on background 
fields with kink solutions.  In fact, this suppression of the $h_u h_d$ 
mass term is unique among the supersymmetry breaking masses of the MSSM 
states: once the gauginos have a large mass, the squark, slepton and 
diagonal Higgs mass terms cannot be protected from low-energy radiative 
corrections, while the Higgsino and off-diagonal Higgs mass terms 
can be.

A simple example accommodating the above mechanism occurs 
in a supersymmetric $SU(5)$ theory in 5D, with the unified 
$SU(5)$ symmetry broken by boundary conditions on the orbifold 
$S^1/Z_2$~\cite{Kawamura:2000ev,Hall:2001pg}.  Supersymmetry may 
be broken on one of the branes by the highest component VEV of a 
chiral superfield $X$.  By localizing $h_d$ towards the brane where 
$X$ resides, while $h_u$ towards the other, we can obtain the pattern 
of the Higgs mass matrix in Eq.~(\ref{eq:H-matrix}).  The quark and 
lepton fields propagate in the bulk, so that the up-type and down-type 
Yukawa couplings arise from the branes where $h_u$ and $h_d$ are 
localized, respectively.  Dangerous dimension six proton decay due 
to gauge boson exchange is also avoided if the matter fields are in 
the bulk because of the split-multiplet structure.  An alternative 
possibility to break supersymmetry is by the $F$-component VEV of 
the radius modulus, or equivalently, through nontrivial boundary 
conditions~\cite{Scherk:1978ta,Barbieri:2001dm}.  The pattern of 
Eq.~(\ref{eq:H-matrix}) can also be obtained in this case, by having 
a similar configuration for the Higgs and matter fields in the 
extra dimension.

\subsection{Models with a single Higgs supermultiplet}
\label{subsec:single-H}

In general, the SM Higgs boson is a linear combination of states at 
the scale $\tilde{m}$.  There is, however, an interesting possibility 
that it comes from a {\it single} supermultiplet in higher dimensions. 
Consider, for example, a supersymmetric $SU(3)_C \times SU(2)_L \times 
U(1)_Y$ gauge theory in 5D, with the extra dimension $y$ compactified 
on $S^1/Z_2$: $0 \leq y \leq \pi R$.  We introduce three generations 
of quark and lepton hypermultiplets $\{M_i, M_i^c \}$ ($M = Q,U,D,L,E$ 
and $i=1,2,3$) and a single Higgs hypermultiplet $\{ H, H^c \}$ in the 
bulk, with the boundary conditions
\begin{equation}
  \left( \begin{array}{c}
    M_i(+,+) \\ M_i^c(-,-) 
  \end{array} \right),
\qquad
  \left( \begin{array}{c}
    H(+,-) \\ H^c(-,+) 
  \end{array} \right).
\label{eq:5D-bc}
\end{equation}
Here, we have denoted a hypermultiplet in terms of two 4D $N=1$ chiral 
superfields, and the first and second signs in parentheses represent 
boundary conditions at $y=0$ and $\pi R$, respectively ($+$ for Neumann 
and $-$ for Dirichlet).  To cancel brane-localized gauge anomalies induced 
by $\{ H, H^c \}$, we also introduce an ``inert Higgs'' hypermultiplet 
$\{ H', H'^c \}$, which has the same boundary conditions but the opposite 
quantum numbers as $\{ H, H^c \}$.  This multiplet, however, does not 
lead to any low energy consequences.

Without supersymmetry breaking, the spectrum of the low energy 
theory consists of 4D $SU(3)_C \times SU(2)_L \times U(1)_Y$ vector 
supermultiplets $V^a$ ($a=1,2,3$) and three generations of quark and 
lepton chiral supermultiplets $Q_i, U_i, D_i, L_i, E_i$.  The KK towers 
of these states have masses $n/R$ ($n=1,2,\cdots$), while those of 
the $H$ and $H'$ hypermultiplets have $(n+1/2)/R$ ($n=0,1,\cdots$). 
We now introduce supersymmetry breaking via the $F$-component VEV 
of the radius modulus, or through nontrivial boundary conditions. 
This shifts the tree-level spectrum of low-lying states as
\begin{equation}
  \left\{ \begin{array}{l}
    m_{A^a_\mu} = 0, \\
    m_{\lambda^a} = \frac{\alpha}{R},
  \end{array} \right.
\qquad
  \left\{ \begin{array}{l}
    m_{q_i,u_i,d_i,l_i,e_i} = 0, \\
    m_{\tilde{q}_i,\tilde{u}_i,\tilde{d}_i,\tilde{l}_i,\tilde{e}_i} 
      = \frac{\alpha}{R},
  \end{array} \right.
\qquad
  \left\{ \begin{array}{l}
    m_{h} = \frac{1/2-\alpha}{R}, \\
    m_{\tilde{h}} = \frac{1}{2R},
  \end{array} \right.
\qquad
  \left\{ \begin{array}{l}
    m_{h'} = \frac{1/2-\alpha}{R}, \\
    m_{\tilde{h}'} = \frac{1}{2R},
  \end{array} \right.
\label{eq:5D-spectrum}
\end{equation}
where $\alpha$ ($0 \leq \alpha \leq 1/2$) is the parameter specifying 
the strength of supersymmetry breaking~\cite{Barbieri:2001dm}, 
and the component fields are defined by $V^a(A_\mu^a,\lambda^a)$, 
$Q_i(\tilde{q}_i,q_i)$ (and similarly for $U_i,D_i,L_i,E_i$), 
$H(h,\tilde{h})$, and $H'(h',\tilde{h}')$.  For $\alpha=1/2$, this 
is essentially the theory of Ref.~\cite{Barbieri:2000vh}.  An important 
difference, however, is that we now take the compactification scale 
$1/R$ to be around the unified scale, rather than at the TeV scale, 
so that the $h$ (and $h'$) states generically obtain masses of order 
$1/4\pi R$ at one loop, which are much larger than the weak scale. 
However, environmental selection can still set $m_h^2$ to be of order 
the weak scale by adjusting various contributions to $m_h^2$ (for 
example by making $\alpha$ deviate slightly from $1/2$ or by introducing 
5D masses for bulk fields; see below).  The low energy particle 
content is then exactly that of the SM:
\begin{equation}
  A_\mu^a,\, q_i,\, u_i,\, d_i,\, l_i,\, e_i,\, h.
\label{eq:5D-SM}
\end{equation}
All the other states decouple at the scale $1/R$.

The Yukawa couplings are obtained by introducing brane-localized operators
\begin{equation}
  S = \int\!d^4x\, dy 
    \left\{ \delta(y)\left[ \frac{(\eta_u)_{ij}}{M_*^{3/2}} 
      Q_i U_j H \right]_{\theta^2} 
    + \delta(y-\pi R)\left[ \frac{(\eta_d)_{ij}}{M_*^{3/2}} Q_i D_j H^c 
      + \frac{(\eta_e)_{ij}}{M_*^{3/2}} L_i E_j H^c \right]_{\theta^2}
    + {\rm h.c.} \right\},
\label{eq:5D-Yukawa-1}
\end{equation}
where $M_*$ is the cutoff scale of the theory, which we take to be 
a factor of a few larger than $1/R$.  The SM Higgs boson, $h(x)$, lies 
in the scalar components of $H$ and $H^c$ as
\begin{equation}
  \left\{ \begin{array}{l}
    h(x,y) = \frac{1}{\sqrt{\pi R}}\, h(x) \cos(m_h y), \\
    h^{c\dagger}(x,y) = -\frac{1}{\sqrt{\pi R}}\, h(x) \sin(m_h y),
  \end{array} \right.
\label{eq:5D-Higgs}
\end{equation}
so that the 4D Yukawa couplings are given by
\begin{equation}
  {\cal L} = (y_u)_{ij} q_i u_j h + (y_d)_{ij} q_i d_j h^\dagger 
    + (y_e)_{ij} l_i e_j h^\dagger,
\label{eq:5D-Yukawa-2}
\end{equation}
with $y_{u,d,e} = (\eta_{u,d,e})_{ij}/(\pi M_* R)^{3/2}$.  Here, we 
have assumed vanishing 5D masses for the bulk hypermultiplets.  The 
form of Eq.~(\ref{eq:5D-Yukawa-2}) is precisely that of the SM.

How does the selection of $m_h^2$ work?  In the limit of $\alpha = 1/2$ 
and vanishing 5D masses, the dominant radiative correction to $m_h^2$ 
comes from top quark/squark loops
\begin{equation}
  \delta m_h^2\bigr|_{\rm top} 
  = -\frac{63 \zeta(3)}{32\pi^4} \frac{y_t^2}{R^2} 
  \simeq -\frac{0.0045}{R^2},
\label{eq:5D-top-loop}
\end{equation}
where we have used $y_t \simeq 0.43$, evaluated at $\approx 
10^{14}~{\rm GeV}$.  Therefore, by making $\alpha$ slightly deviate 
from $1/2$
\begin{equation}
  \alpha \simeq \frac{1}{2} - \sqrt{-\delta m_h^2\bigr|_{\rm top} R^2} 
  \simeq 0.43,
\label{eq:5D-alpha}
\end{equation}
we can set $m_h^2$ to have the correct, weak scale (and negative) value.%
\footnote{The precise value of $\alpha$ would be changed by the existence 
 of brane-localized terms, such as $\delta(y)[H H']_{\theta^2}$, but 
 our basic conclusion does not change.  Below we assume that these 
 terms are absent.}
Alternatively, we may introduce 5D bulk masses for top 
hypermultiplets.  In this case the top-loop contribution of 
Eq.~(\ref{eq:5D-top-loop}) is suppressed~\cite{Barbieri:2002sw}, 
so that it can be canceled with the gauge loop contribution
\begin{equation}
  \delta m_h^2\bigr|_{\rm gauge} 
  = \frac{7 \zeta(3)}{64\pi^4} \frac{3g^2 + g'^2}{R^2} 
  \simeq \frac{0.0014}{R^2},
\label{eq:5D-gauge-loop}
\end{equation}
even for $\alpha = 1/2$, leaving the correct value for $m_h^2$.

An interesting property of the theory considered here is that the 
tree-level Higgs quartic coupling is given by
\begin{equation}
  \lambda = \frac{g^2 + g'^2}{8},
\label{eq:5D-lambda}
\end{equation}
{\it regardless of the value of $\alpha$}---there is no free parameter 
such as $\beta$ in 4D supersymmetric theories.  This is a consequence 
of the $SU(2)_R$ symmetry and the fact that the SM Higgs boson 
resides in a single higher dimensional supermultiplet.  Therefore, 
at the leading order, {\it the theory just below $1/R$ is precisely 
the SM but with the Higgs quartic coupling constrained as in 
Eq.~(\ref{eq:5D-lambda})}.  The relation of Eq.~(\ref{eq:5D-lambda}) 
can receive corrections from brane-localized kinetic terms.  These 
effects are suppressed by the volume factor (and possibly also by 
a loop factor), which we estimate to give an $O(10\%)$ correction 
to $\lambda$.  This is translated into an uncertainty of the Higgs 
mass prediction at the level of a GeV.

It is straightforward to construct unified models along the lines 
discussed here.  For example, we can consider a supersymmetric $SU(5)$ 
theory in 6D with $SU(5)$ broken along one extra dimension while 
supersymmetry along the other.  For $\alpha \neq 1/2$, we can even 
use the same dimension to break both supersymmetry and a unified symmetry. 
We simply need to embed the model discussed above into $SU(5)$, and 
break $SU(5)$ by boundary conditions at $y = \pi R$ (and supersymmetry 
by Eq.~(\ref{eq:5D-alpha})).  In this theory, some of the unified states 
have a tree-level mass of $(1/2-\alpha)/R$ and thus lighter than $1/R$ 
by about an order of magnitude, and the colored triplet Higgsinos 
obtain their masses through brane-localized operators.  Unification 
of the SM gauge couplings receives corrections both from KK towers and 
brane-localized gauge kinetic operators.  The deviation from single-scale 
exact unification in the SM may arise from these corrections.

\section{Evidence for the Multiverse from the Higgs Boson Mass}
\label{sec:discussion}

The Standard Model is remarkably successful, correctly predicting the 
results of three decades of particle physics experiments at both the 
high energy and high precision frontiers.  From the absence of proton 
decay, to precision measurements of the electroweak sector, to rare 
quark and lepton flavor violation and even $CP$ violation, the SM has 
consistently and repeatedly passed every experimental challenge.  Indeed, 
the electroweak and flavor data now constrain new physics at the TeV scale 
so strongly, that the resulting difficulties in developing alternative 
natural theories have become a main focus of much research.  Why then 
do we resist the simplest possibility, that {\it the SM is the correct 
description of nature up to unified energy scales}?  This question 
seems particularly pressing since the SM, valid to very high energies, 
predicts $110~{\rm GeV} \simlt M_H \simlt 190~{\rm GeV}$, precisely 
the range selected by limits from direct searches and from precision 
electroweak data.

There are two key deficiencies of the SM, one theoretical and one 
observational.  On the theoretical side, the lack of naturalness of 
the Higgs mass parameter has been the essential driving force for 
a variety of extensions of the theory at the TeV scale.  However, 
the cosmological constant is a numerically more severe fine-tuning 
problem, and has no known symmetry solution.  The realization that 
this problem has an environmental solution~\cite{Weinberg:1987dv} 
motivated the discovery of a possible environmental understanding 
for the weak scale~\cite{Agrawal:1997gf}.  The discovery of dark 
energy~\cite{Perlmutter:1998np} provided remarkable evidence for 
environmental selection: dark energy with $w = -1$ is a necessary 
consequence of the environmental solution of the cosmological constant 
problem, and requires no physics beneath unified scales beyond the 
SM and general relativity.  The absence of dark energy would have 
demonstrated that environmental selection had failed its greatest 
opportunity.  Of course, an enormous landscape of vacua is required, 
as well as a cosmological mechanism for populating these vacua to form 
a multiverse.  The realization that string theory~\cite{Bousso:2000xa} 
and eternal inflation~\cite{Guth:1982pn} may yield such a multiverse, 
opens the door to a firm theoretical foundation for the environmental 
selection of both the cosmological constant and the weak scale.

Dark matter provides the other key deficiency of the SM, but it is 
a theoretical extrapolation to attribute this DM to particles with 
weak scale mass.  Even if DM is composed of cold particles, nothing 
is known observationally about their mass.  The WIMP hypothesis provides 
an intriguing possibility that the abundance of DM may be derived from 
the weak scale, but is subject to uncertainties of several orders of 
magnitude.  If the SM is valid to unified scales, the most compelling 
candidate for DM is axions.  The strong $CP$ problem requires a symmetry 
solution, since there is no environmental need for low $\bar{\theta}$. 
The axion solution, theoretically motivated by string theory, cannot be 
implemented at the weak scale, and requires $f_A \simgt 10^9~{\rm GeV}$. 
Even if $f_A$ is as large as the unified scale, environmental selection 
can act on the initial axion misalignment angle to avoid overproduction 
of DM~\cite{Linde:1987bx}.

Over more than three decades, much effort has been expended on extensions 
of the SM at the TeV scale.  Is there any experimental evidence that any 
of these alternatives are to be preferred over the SM?  While there is 
no direct experimental evidence for any such extension, in the case of 
weak scale supersymmetry gauge coupling unification occurs with greater 
precision than in the SM.  When first discovered at LEP, this result 
appeared highly significant.  Precise data outweighed the well-known 
cosmological and flavor problems of supersymmetry, which received renewed 
attention.  However, the LEP2 limit on the Higgs boson mass provided 
contrary data, that imposed a precise numerical naturalness problem on 
supersymmetry.  Is the reduction of the unified threshold corrections 
on gauge coupling unification by an order of magnitude worth the required 
fine-tuning of the theory at the percent level?

With environmental selection on a multiverse, the minimal effective 
theory below the unified scale, SM~$+$~GR, has no deficiencies.  Instead 
of introducing problems by augmenting the SM at the TeV scale, it seems 
worthwhile seeking additional evidence for environmental selection in 
the minimal effective theory.  In this paper we presented a precise 
and robust prediction for the Higgs boson mass.  We argued that a 
supersymmetric boundary condition on the Higgs quartic coupling is likely, 
yielding a Higgs boson mass range of $(128~\mbox{--}~141)~{\rm GeV}$. 
The upper edge of $141~{\rm GeV}$ is particularly interesting, arising 
from the special situation that the SM Higgs boson lies dominantly 
in a single supermultiplet as occurs, for example, with an approximate 
Peccei-Quinn symmetry.  Corrections at the supersymmetry breaking 
scale $\tilde{m}$ are remarkably small: $0.1~\mbox{--}~0.3~{\rm GeV}$ 
from top squark loops and $0.3~{\rm GeV}$ from varying $\tilde{m}$ 
by two orders of magnitude.  The dominant uncertainty in the prediction, 
of $\pm 2~{\rm GeV}$, arises from the present uncertainties in $m_t$ 
and $\alpha_s$, but measurements at future collider experiments could 
reduce this to $\pm 0.3~{\rm GeV}$, so that the prediction could be 
tested down to the level of $0.4\%$.

Going beyond this minimal scenario, there are several physical origins 
of corrections in the GeV region.  If neutrino masses arise from the 
seesaw mechanism, the corrections to the Higgs boson mass are negligible 
except, in a certain region of parameter space with $\tilde{m} \gg 
10^{14}~{\rm GeV}$, the Higgs mass could be raised by about a GeV.%
\footnote{This implies that leptogenesis~\cite{Fukugita:1986hr} can be 
 accommodated without affecting the Higgs mass prediction.}
Higher dimensional theories having a single Higgs supermultiplet lead 
to the Higgs mass being near the upper edge of $141~{\rm GeV}$, but 
brane-localized kinetic terms lead to uncertainties of about a GeV. 
Finally, while adding states at the weak scale beyond those of the SM 
typically destroys the prediction, there are a few minimal cases that 
yield mild perturbations; for example, a single vector-like lepton 
increases the Higgs boson mass only by $0.35~{\rm GeV}$.

\begin{figure}[t]
  \center{\includegraphics[scale=0.77]{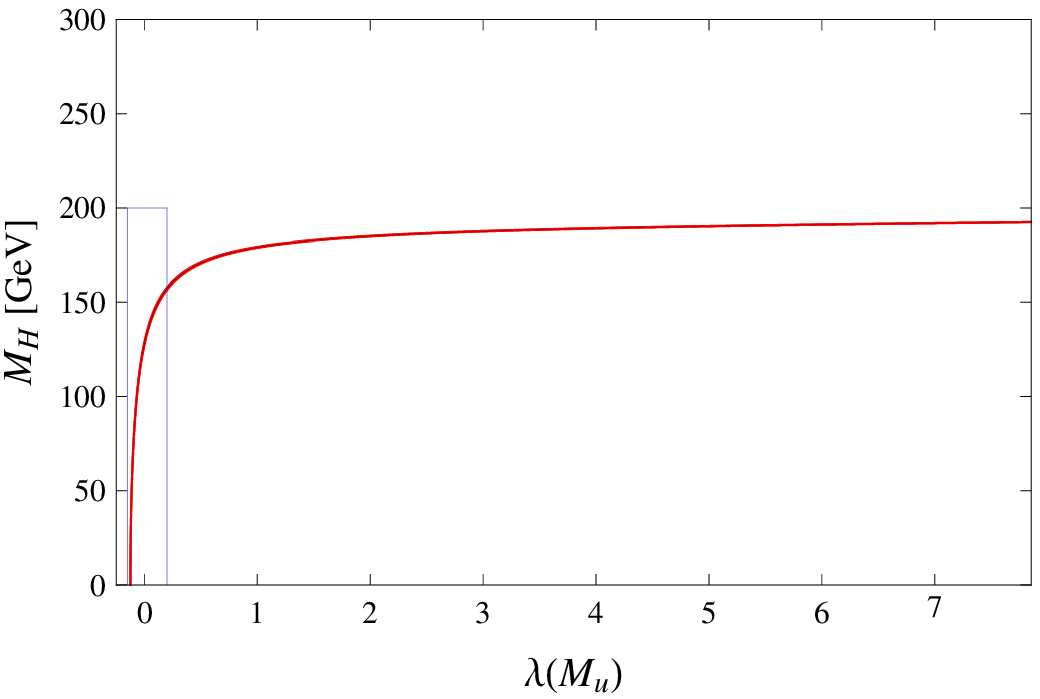}
  \hspace{0.5cm}
  \includegraphics[scale=0.77]{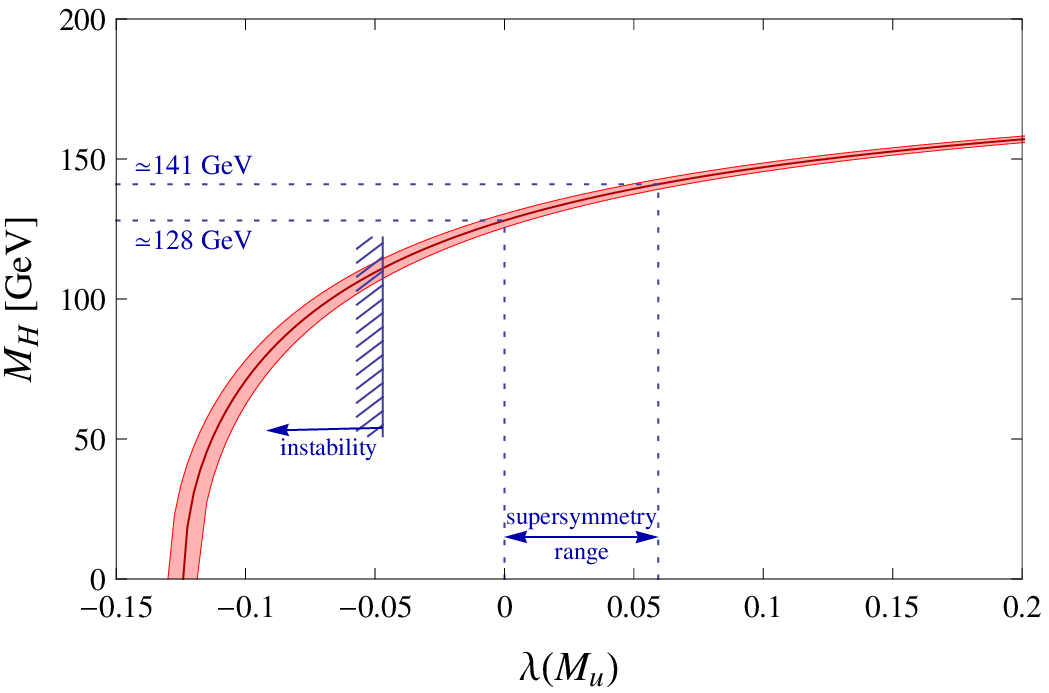}}
\caption{The Higgs boson mass as a function of $\lambda(M_u)$ for 
 the SM valid up to $M_u$, with a wide range of $\lambda(M_u)$ in 
 the left panel and an expansion of the region of small $\lambda(M_u)$ 
 in the right panel.  The values of $M_u$ and $\alpha_s$ are fixed at 
 $M_u = 10^{14}~{\rm GeV}$ and $\alpha_s(M_Z) = 0.1176$, respectively, 
 and the shaded bands represent the variation of the Higgs boson mass 
 for $m_t = 173.1 \pm 1.3~{\rm GeV}$.  For large $\lambda(M_u)$, 
 the left panel shows that the Higgs boson mass asymptotes to about 
 $190~{\rm GeV}$.  The right panel shows the supersymmetric range 
 of $\lambda(M_u)$, with a corresponding Higgs boson mass range of 
 $(128~\mbox{--}~141)~{\rm GeV}$, as well as the electroweak vacuum 
 stability bound of $\lambda(M_u) \simgt -0.05$.}
\label{fig:SM-Higgs}
\end{figure}
Are there other special values for the Higgs boson mass that would 
provide evidence for the multiverse?  In Figure~\ref{fig:SM-Higgs} 
we show the Higgs mass as a function of the quartic coupling at the 
unified scale $M_u$, assuming only that the effective theory below 
$M_u$ is the SM.  The left panel gives a wide range of $\lambda(M_u)$, 
while the right panel expands the region of small $\lambda(M_u)$. 
We draw attention to four special values of the Higgs mass:%
\footnote{To simplify the presentation, we take the scale at which the 
 quartic coupling takes special values to be $M_u$.  In fact, depending 
 on the case, this scale could be $\tilde{m}$ or $M_*$, but we do not 
 expect these scales to differ by many orders of magnitude.}
\begin{itemize}
\item
$M_H \sim 190~{\rm GeV}$:
results from a very wide range of $\lambda(M_u) \simgt 2$, including 
the case of strong coupling, $\lambda(M_u) \approx 2\pi$.
\item
$M_H \simeq 141~{\rm GeV}$:
results from the supersymmetric boundary condition $\lambda(M_u) = 
\{g^2(M_u) + g'^2(M_u)\}/8$, as explored in detail in this paper.
\item
$M_H \simeq 128~{\rm GeV}$:
results from $\lambda(M_u) = 0$.
\item
$M_H \simeq 112~{\rm GeV}$:
this is the smallest Higgs boson mass theoretically allowed, since 
smaller values would lead to cosmological instabilities in the 
electroweak vacuum.  A value close to this may result from a multiverse 
distribution function that is peaked strongly towards large and 
negative $\lambda(M_u)$~\cite{Feldstein:2006ce}.
\end{itemize}

Since $M_u$ is not well determined by gauge coupling unification, 
an important question is the sensitivity of these four special Higgs 
mass values to variations in $M_u$.  In the first three cases the 
sensitivity depends on how close the RG trajectory is to the quasi-fixed 
point trajectory.  The case of strong coupling is very far from the 
fixed point and has significant sensitivity, with the Higgs mass 
varying by $\pm 10~{\rm GeV}$ for $M_u = 10^{14 \pm 2}~{\rm GeV}$. 
A Higgs mass in this range would be indicative of a multiverse that 
has a high probability for a large quartic coupling, but the evidence 
would be rather weak.  The cases of $\lambda(M_u) = \{g^2(M_u) 
+ g'^2(M_u)\}/8$ and $\lambda(M_u) = 0$ are much closer to the 
quasi-fixed point, giving Higgs mass variations of only $\pm 
0.3~{\rm GeV}$ and $\pm 1.0~{\rm GeV}$, respectively, for the same 
variation in $M_u$.  Thus a Higgs mass near $128~{\rm GeV}$ would 
provide strong evidence for the multiverse, although not quite as 
strong as might occur for a value near $141~{\rm GeV}$.  The case 
of the smallest Higgs mass is more complicated, since it involves 
tunneling, but it is also insensitive to variations in $M_u$.  Thus 
a value of the Higgs mass very close to the minimal value would 
also yield evidence for the multiverse, although for this to occur 
requires a very sharp variation in the multiverse probability 
distribution for $\lambda(M_u)$.

Although this paper has focused on the Higgs boson mass near 
$141~{\rm GeV}$, a value near $128~{\rm GeV}$ is also very interesting. 
These two values are the upper and lower edge values allowed by the 
supersymmetric boundary condition of Eq.~(\ref{eq:lambda-bc_beta}), 
corresponding to $\beta = 0~\mbox{or}~\pi/2$ and $\beta = \pi/4$ 
respectively.  Studying the mass matrix for the two Higgs doublets 
in the supersymmetric theory, the former occurs when a diagonal entry 
is much larger than the off-diagonal entry, while the latter occurs 
if the off-diagonal entry is much larger than the splitting between 
the diagonal entries, as would occur if the mass matrix were invariant 
under a symmetry that interchanged the two doublets.  Our discussions 
of the corrections to the Higgs mass for the large $\tan\beta$ case 
apply also to the case of $\tan\beta$ near unity except, as noted 
above, the convergence effect from the quasi-fixed point of the 
quartic coupling is not quite as strong.  For example, the top squark 
loops at $\tilde{m}$ lead to an uncertainty in the Higgs mass of 
$0.2~{\rm GeV}$ for $A_t = m_{\tilde{t}}$.  Also the uncertainty 
in the Higgs mass arising from the present experimental uncertainties 
on $m_t$ and $\alpha_s$ is $\pm 3~{\rm GeV}$, $50\%$ larger than at 
the $141~{\rm GeV}$ edge.  Finally we should note that a Higgs mass 
near $128~{\rm GeV}$ occurs in any theory where the SM Higgs doublet 
is a pseudo Nambu-Goldstone boson, with a vanishing tree-level 
potential at $M_u$.  An example of this occurs when the Higgs boson 
is identified as an extra-dimensional component of a gauge field 
in a non-supersymmetric 5D theory~\cite{Gogoladze:2007qm}.

Much of the excitement in particle physics in the coming decade will follow 
from unraveling the origin of the weak scale.  Three clear options are
\begin{itemize}
\item
{\it Weak scale supersymmetry.} This will confirm the indirect evidence 
of gauge coupling unification, and allow many measurements that provide 
a window to much higher energy scales.
\item
{\it New strong dynamics.} A composite Higgs, or even a Higgsless 
theory, would make the TeV scale extremely rich, and may even herald 
new spatial dimensions.
\item
{\it Environmental selection.} Precision measurements of SM parameters 
may point to a multiverse and the need for a clearer understanding of 
the catastrophic boundaries at which selection takes place.
\end{itemize}
Strong evidence for the multiverse would result if the LHC discovered 
a Higgs boson mass close to $141~{\rm GeV}$, or $128~{\rm GeV}$, and 
no new physics beyond the SM.  This would add greatly to the evidence 
from the cosmological constant problem and the discovery of dark energy. 
The two fine-tuning problems of SM~$+$~GR would have a common solution, 
with other solutions either unknown or disproved.  Through nuclear 
stability, the multiverse accounts for the values of the up quark, 
down quark and electron masses remarkably well~\cite{Hall:2007ja}. 
Furthermore, the multiverse may also explain the cosmological mystery 
of why the time scales of structure formation, galaxy cooling and 
vacuum domination do not differ by many orders of magnitude, but are 
all comparable to the present age of the universe~\cite{Bousso:2009ks}. 
Instead of discovering more symmetries, the LHC may play a key part 
in the accumulation of evidence for more universes.

\section*{Acknowledgments}

We thank Gilly Elor and Piyush Kumar for useful discussions.  This work 
was supported in part by the Director, Office of Science, Office of 
High Energy and Nuclear Physics, of the US Department of Energy under 
Contract DE-AC02-05CH11231.  The work of L.H. was supported in part 
by the National Science Foundation under grant PHY-0457315, and that 
of Y.N. under grants PHY-0555661 and PHY-0855653.

\appendix

\section{Supersymmetric Threshold Corrections at {\boldmath $\tilde{m}$}}
\label{app:thres}

The leading-log corrections to the supersymmetric boundary condition, 
$\lambda = (g^2 + g'^2)/8$, when matching between the SM and a theory 
with the states of the MSSM, at a scale $\tilde{m}$, are
\begin{eqnarray}
  \delta_{\rm LL} 
  &=& \frac{1}{32\pi^2 \lambda} \biggl( 
    12 y_t^4 \ln\frac{m_{\tilde{Q}_3}^{\frac{1}{2}} 
        m_{\tilde{U}_3}^{\frac{1}{2}}}{\tilde{m}} 
    - 3 g_2^2 y_t^2 \ln\frac{m_{\tilde{Q}_3}}{\tilde{m}} 
    - \frac{9}{5} g_1^2 y_t^2 
      \ln\frac{m_{\tilde{Q}_3}^{-\frac{1}{3}} 
        m_{\tilde{U}_3}^{\frac{4}{3}}}{\tilde{m}}
\nonumber\\
  && {} \qquad\quad - \frac{7}{3} g_2^4 
      \ln\frac{m_\chi^{\frac{3}{2}} m_\lambda^{\frac{2}{7}} 
        m_{\tilde{h}}^{\frac{1}{7}} m_{\tilde{Q}}^{-\frac{9}{14}} 
        m_{\tilde{L}}^{-\frac{3}{14}} m_H^{-\frac{1}{14}}}{\tilde{m}}
\nonumber\\
  && {} \qquad\quad + \frac{24}{25} g_1^4 
      \ln\frac{m_\chi^{-\frac{3}{16}} m_{\tilde{h}}^{-\frac{1}{8}} 
        m_{\tilde{Q}}^{\frac{1}{16}} m_{\tilde{U}}^{\frac{1}{2}} 
        m_{\tilde{D}}^{\frac{1}{8}} m_{\tilde{L}}^{\frac{3}{16}} 
        m_{\tilde{E}}^{\frac{3}{8}} m_H^{\frac{1}{16}}}{\tilde{m}} 
    \biggr),
\label{eq:delta-LL-full}
\end{eqnarray}
where $m_\lambda$, $m_{\tilde{h}}$, $m_{\tilde{Q}_i, \tilde{U}_i,
\tilde{D}_i, \tilde{L}_i, \tilde{E}_i}$ ($i=1,2,3$), and $m_H$ are 
the gaugino, Higgsino, squark and slepton, and heavy Higgs boson 
masses, $m_\chi \equiv {\rm max}\{ m_\lambda, m_{\tilde{h}} \}$, 
and $m_{\tilde{\Phi}} \equiv (m_{\tilde{\Phi}_1} m_{\tilde{\Phi}_2} 
m_{\tilde{\Phi}_3})^{1/3}$ ($\Phi = Q,U,D,L,E$).  Here, we have taken 
the wino and bino masses to be equal, $m_\lambda$, which is generically 
a good approximation since $\tilde{m}$ is not far from $M_u$.  The 
dependence on the matching scale $\tilde{m}$ cancels that from the 
RG scaling in the SM, given by (\ref{eq:delta-MH-2}).

Since $\tilde{m} \sim M_u$, it is appropriate to make an approximation 
$g_1 = g_2 \equiv g_u$, leading to
\begin{eqnarray}
  \delta_{\rm LL} 
  &=& \frac{1}{32\pi^2 \lambda} \biggl\{ 
    \Bigl(12 y_t^4 - \frac{24}{5} g_u^2 y_t^2 \Bigr) 
      \ln\frac{m_{\tilde{Q}_3}^{\frac{1}{2}} 
        m_{\tilde{U}_3}^{\frac{1}{2}}}{\tilde{m}} 
\nonumber\\
  && {} - \frac{103}{75} g_u^4 
      \ln\frac{m_\chi^{\frac{276}{103}} m_\lambda^{\frac{50}{103}} 
        m_{\tilde{h}}^{\frac{34}{103}} m_{\tilde{Q}}^{-\frac{117}{103}} 
        m_{\tilde{U}}^{-\frac{36}{103}} m_{\tilde{D}}^{-\frac{9}{103}} 
        m_{\tilde{L}}^{-\frac{51}{103}} m_{\tilde{E}}^{-\frac{27}{103}} 
        m_H^{-\frac{17}{103}}}{\tilde{m}} 
    \biggr\}.
\label{eq:delta-LL-unif}
\end{eqnarray}
As discussed in section~\ref{subsec:Higgs-SM}, a useful choice 
of $\tilde{m}$ is the one that makes $\delta_{\rm LL}$ vanish. 
This scale can be estimated by assuming that the second line of 
Eq.~(\ref{eq:delta-LL-unif}) is dominated by the gaugino piece:
\begin{equation}
  \delta_{\rm LL} 
  \simeq \frac{1}{32\pi^2 \lambda} \biggl\{ 
    \Bigl(12 y_t^4 - \frac{24}{5} g_u^2 y_t^2 \Bigr) 
      \ln\frac{m_{\tilde{t}}}{\tilde{m}} 
    - \frac{326}{75} g_u^4 \ln\frac{m_\lambda}{\tilde{m}}
    \biggr\},
\label{eq:delta-LL-approx}
\end{equation}
where $m_{\tilde{t}} \equiv (m_{\tilde{Q}_3} m_{\tilde{U}_3})^{1/2}$, 
and we have taken $m_{\tilde{h}} < m_\lambda$.  The logarithmic terms for 
each particle dropped from the second line of Eq.~(\ref{eq:delta-LL-unif}) 
have coefficients that are smaller than in the gaugino term by a factor 
of $8$ or more.  In fact, a random deviation of these superparticle masses 
from $\tilde{m}$ by a similar amount to $m_{\tilde{t}}$ and $m_\lambda$ 
will not contribute to $\delta_{\rm LL}$ as much as the terms shown 
in Eq.~(\ref{eq:delta-LL-approx}).  By equating the expression of 
Eq.~(\ref{eq:delta-LL-approx}) to zero, we obtain
\begin{equation}
  \tilde{m} \simeq \left( \frac{m_\lambda^{163 g_u^4}} 
    {m_{\tilde{t}}^{450 y_t^4 - 180 g_u^2 y_t^2}} 
    \right)^{\frac{1}{163 g_u^4 - 450 y_t^4 + 180 g_u^2 y_t^2}} 
  \simeq \frac{m_\lambda^{1.6}}{m_{\tilde{t}}^{0.6}}.
\label{eq:app-mtilde}
\end{equation}
This is the expression quoted in Eq.~(\ref{eq:mtilde}).

\end{document}